\documentclass[conference]{IEEEtran}

\usepackage{etex}

\newcommand{\maciej}[1]{\textcolor{blue}{[Maciej: #1]}}

\newcommand{\harun}[1]{\textcolor{olive}{#1}}
\newcommand{\haruncor}[2]{\sout{#1}\textcolor{olive}{#2}}
\newcommand{\haruncomment}[1]{\textcolor{olive}{[Harun: #1]}}
\newcommand{\revision}[1]{\noindent\textcolor{red}{[#1]}}

\usepackage[font={sf,scriptsize}]{caption}
\usepackage{subfig}

\usepackage{makecell,cellspace}
\usepackage{graphicx}
\graphicspath{{figs/},{plots/}}
\usepackage{booktabs}
\usepackage{epstopdf}
\usepackage{url}
\usepackage{placeins}
\usepackage{amsmath,amssymb,amsfonts,amsthm}
\usepackage{mathtools,mathrsfs}
\usepackage{multirow}
\usepackage{rotating}
\usepackage{pbox}
\usepackage[normalem]{ulem}

\usepackage{inconsolata}
\usepackage{listings}

\usepackage{etoolbox}

\makeatletter
\patchcmd{\ttlh@hang}{\parindent\z@}{\parindent\z@\leavevmode}{}{}
\patchcmd{\ttlh@hang}{\noindent}{}{}{}
\makeatother

\usepackage[utf8]{inputenc}
\newcommand{\cref}[1]{\ref{#1}}

\usepackage[10pt]{moresize}

\usepackage{enumitem}

\usepackage{xcolor}

\definecolor{darkgrey}{RGB}{140,140,140}
\definecolor{lightgrey}{RGB}{200,200,200}

\usepackage{scalerel,stackengine}
\stackMath
\newcommand\rwh[1]{%
\savestack{\tmpbox}{\stretchto{%
  \scaleto{%
      \scalerel*[\widthof{\ensuremath{#1}}]{\kern-.6pt\bigwedge\kern-.6pt}%
          {\rule[-\textheight/2]{1ex}{\textheight}}%WIDTH-LIMITED BIG WEDGE
            }{\textheight}% 
}{0.5ex}}%
\stackon[1pt]{#1}{\tmpbox}%
}

\usepackage[normalem]{ulem}

\usepackage{bm}
\newcommand{\vcr}[1]{\bm{#1}}
\newcommand{\mat}[1]{\bm{#1}}

\usepackage{amsmath,amssymb,amsfonts}
\usepackage{graphicx}
\usepackage{textcomp}
\def\BibTeX{{\rm B\kern-.05em{\sc i\kern-.025em b}\kern-.08em
    T\kern-.1667em\lower.7ex\hbox{E}\kern-.125emX}}

\usepackage{fontawesome}
\usepackage{pifont}
\usepackage{textcomp}

\usepackage{soul}

\usepackage{dblfloatfix}

\usepackage{mathrsfs}
\usepackage{amsmath}

\usepackage{algorithm,algorithmicx,algpseudocode}

\usepackage{tikz}
\usetikzlibrary{tikzmark}

\usepackage{xpatch}
\expandafter\xpatchcmd
\csname pgfk@/tikz/every picture/.@cmd\endcsname
{\thepage}{\arabic{page}}{}{}

\tikzstyle{comment} = [draw, fill=blue!70, text=white, text width=3cm, minimum height=1cm, rounded corners, align=left, font=\scriptsize]
\tikzstyle{background_alg} = [draw, fill=blue!20, opacity=0.4, inner sep=4pt, rounded corners=2pt]

\definecolor{vlgray}{rgb}{0.77 0.77 0.77}
\definecolor{ablack}{rgb}{0.2 0.2 0.2}
\definecolor{grayblack}{rgb}{0.4 0.4 0.4}

\usepackage[customcolors]{hf-tikz}
\hfsetbordercolor{white}
\hfsetfillcolor{vlgray}

\newcounter{highlight}

\newcounter{Ahighlight}

\newif\ifrev
\revfalse

\sethlcolor{white}

\begin{document}

%\title{\vspace{-0.5em}Scaling Algebraic Jaccard similarity Computations\\for Fast and Accurate Genome Distance Derivation\vspace{-0.25em}}
%
%\title{\vspace{-0.5em}Fast and Accurate Genome Distance Derivations\\with Communication-Efficient Matrix Multiplication\vspace{-0.25em}}
%
%\title{High-Performance Distributed Jaccard similarity\\with Communication-Efficient Matrix Multiplication}

\title{Communication-Efficient Jaccard similarity for\\\hspace{-0.5em}High-Performance Distributed Genome Comparisons}

\author{\IEEEauthorblockN{Maciej Besta$^1$$^\dagger$, Raghavendra Kanakagiri$^5$$^\dagger$, Harun Mustafa$^{1,3,4}$, Mikhail Karasikov$^{1,3,4}$,\\Gunnar R\"atsch$^{1,3,4}$, Torsten Hoefler$^1$, Edgar Solomonik$^2$}
\IEEEauthorblockA{
\textit{$^1$Department of Computer Science, ETH Zurich}\\
\textit{$^2$Department of Computer Science, University of Illinois at Urbana-Champaign}\\
\textit{$^3$University Hospital Zurich, Biomedical Informatics Research}\\
\textit{$^4$SIB Swiss Institute of Bioinformatics} \\
\textit{$^5$Department of Computer Science, Indian Institute of Technology Tirupati
%
%\vspace{-0.25em}
%
} \\
}
$^\dagger$Both authors contributed equally to this work.
}

\maketitle

\begin{abstract}
The Jaccard similarity index is an important measure of the overlap of two
sets, widely used in machine learning, computational genomics, information
retrieval, and many other areas. 
We design and implement
SimilarityAtScale, the first communication-efficient distributed algorithm for
computing the Jaccard similarity among pairs of large datasets.
Our algorithm provides an efficient encoding of this problem into a multiplication of sparse matrices.
Both the encoding and sparse matrix product are performed in a way that minimizes data movement in terms of communication and synchronization costs.
We apply our algorithm to obtain similarity among all pairs of a set of large samples of genomes.
This task is a key part of modern metagenomics analysis and an evergrowing need due to the increasing 
availability of high-throughput DNA sequencing data. The resulting scheme is the first to enable accurate Jaccard
distance derivations for massive datasets, using large-scale distributed-memory
systems. We package our routines in a tool, called GenomeAtScale, that combines
the proposed algorithm with tools for processing input sequences.
Our evaluation on real data illustrates that one can use
GenomeAtScale to 
effectively employ tens of thousands of processors to reach new frontiers in
large-scale genomic and metagenomic analysis.
While
GenomeAtScale can be used to foster DNA research,
the more general
underlying SimilarityAtScale algorithm may be used for high-performance distributed
similarity computations in 
other data analytics application domains.

\end{abstract}

\begin{IEEEkeywords}
Distributed Jaccard Distance,
Distributed Jaccard similarity,
Genome Sequence Distance, Metagenome Sequence Distance,
High-Performance Genome Processing,
k-Mers, Matrix-Matrix Multiplication,
Cyclops Tensor Framework
\end{IEEEkeywords}

{\noindent\small\textbf{Code and data:} \url{https://github.com/cyclops-community/jaccard-ctf}}

\section{Introduction}
\label{sec:intro}

The concept of \emph{similarity} 
has
gained much attention in different areas of
data analytics~\cite{tan2018introduction}. A popular method of assessing a
similarity of two entities is based on computing their Jaccard
\emph{similarity} index $J(A, B)$~\cite{jaccard1901etude,
levandowsky1971distance}. $J(A,B)$ is a statistic that assesses the overlap of
two sets $A$ and $B$. It is defined as the ratio between the size of the
intersection of $A$ and $B$ and the size of the union of $A$ and $B$: $J(A, B)
= |A \cap B| / |A \cup B|$. A closely related notion, the Jaccard
\emph{distance}~$d_J = 1 - J$, measures the \emph{dissimilarity} of sets. The
general nature and simplicity of the Jaccard similarity and distance 
has allowed for their wide use in numerous domains, for example computational genomics (comparing
DNA sequencing data sets), machine learning (clustering, object recognition)~\cite{ben2019modular},
information retrieval (plagiarism detection), and many
others~\cite{theodoridis2009pattern, dong2006hierarchical,
rezatofighi2019generalized, schaeffer2007graph, selivanov2016text2vec,
zielezinski2017alignment, strehl2000impact}.

We focus on the application of the Jaccard similarity index and distance to
genetic distances between high-throughput DNA sequencing samples, a problem of
high importance for different areas of computational
biology~\cite{seth2014exploration,zielezinski2017alignment,choi2018libra}.
Genetic distances are frequently used to approximate the evolutionary distances
between the species or populations represented in sequence sets in a
computationally intractable
manner~\cite{seth2014exploration,ondov2016mash,zielezinski2017alignment,choi2018libra}.
This enables or facilitates analyzing the evolutionary history of populations
and species~\cite{ondov2016mash}, the investigation of the biodiversity of a
sample, as well as other
applications~\cite{zielezinski2017alignment,popic2018fast}.  However, the
enormous sizes of today's high-throughput sequencing datasets often make it
infeasible to compute the exact values of $J$ or
$d_J$~\cite{stephens2015biggenomical}.  Recent works \hl{(such as
Mash~}\cite{ondov2016mash}) propose approximations, for example using the
MinHash representation of $J(A,B)$, which is the primary locality-sensitive
hashing (LSH) scheme used for genetic
comparisons~\cite{kucherov2019evolution}. Yet, these
approximations often lead to inaccurate approximations of $d_J$ for highly
similar pairs of sequence sets, and tend to be ineffective for computation of a
distance between highly dissimilar sets unless very large sketch sizes are
used~\cite{ondov2016mash}, as noted by the Mash
authors~\cite{ondov2016mash}.
Thus, developing a \emph{scalable}, \emph{high-performance}, and
\emph{accurate} scheme for computing the Jaccard similarity index is an open
problem of considerable relevance for genomics computations and numerous other
applications in general data analytics.

\ifrev

\marginpar{\vspace{-11em}\colorbox{orange}{\textbf{R4}}}

\marginpar{\vspace{6em}\colorbox{orange}{\textbf{R4}}}

\marginpar{\vspace{2em}\colorbox{orange}{\textbf{R1}}}

\fi

Yet, there is little research on high-performance distributed
derivations of either $J$ or $d_J$. Existing works target very small
datasets (e.g., 16MB of raw input text~\cite{cosulschi2015scaling}),
only provide simulated evaluation for experimental
hardware~\cite{krawezik2018implementing}, focus on inefficient MapReduce
solutions~\cite{burkhardt2014asking, bank2008calculating,vernica2010efficient}
that need asymptotically more communication due to using the
allreduce collective communication pattern over reducers~\cite{hoefler2014energy}, are limited to a
single server~\cite{fender2017parallel, sachdeva2009evaluating,
scripps2015parallelizing},
use an approach based on deriving the Cartesian product, which may require
quadratic space (infeasible for the large input datasets considered in this work)~\cite{hu2019output},
or do not target parallelism or distribution at all~\cite{bayardo2007scaling}.
Most works target novel use cases of $J$ and $d_J$~\cite{binanto2018comparison,
brekhna2019experimental, valari2013continuous, kogge2016jaccard}, or
mathematical foundations of these measures~\cite{kosub2019note,
moulton2018maximally}. We attribute this to two factors. First, many domains
(for example genomics) only recently discovered the usefulness of Jaccard
measures~\cite{zielezinski2017alignment}. 
Second, the rapid growth 
%
\iffalse
(faster than Moore's Law~\cite{stephens2015biggenomical}) 
\fi
%
of the availability
of high-throughput sequencing data and its increasing relevance to genetic analysis
have meant that previous complex genetic analysis methods are falling out of favor
due to their poor scaling properties~\cite{zielezinski2017alignment}.
\emph{To the best of our knowledge, no work provides a scalable
high-performance solution for
computing
$J$ or $d_J$.}
%\revision{I believe this is still true? Though BELLA can be used for jaccard similarity index calculation, the solution is not scalable.}

In this work, we design and implement \textbf{SimilarityAtScale}: the first
algorithm for distributed, fast, and scalable derivation of the Jaccard
Similarity index and Jaccard distance.  
{We follow~\cite{bella2019}, devising an algorithm based on an algebraic
formulation of Jaccard similarity matrix computation as sparse matrix multiplication.
%The key idea behind our algorithm is to
%\emph{model the problem algebraically}, as a series of matrix operations, and
Our main contribution is a \emph{communication-avoiding algorithm} for both preprocessing
the sparse matrices as well as computing Jaccard similarity in parallel via sparse matrix multiplication.}
We
use our algorithm as a core element of \textbf{GenomeAtScale}, a tool
that we develop to facilitate high-performance genetic distance computation.
GenomeAtScale enables \emph{the first 
massively-parallel and provably accurate
calculations of Jaccard distances
%
\iffalse
%
\haruncomment{we should be careful with wording here. in the end, all of these are meant to be tractable
approximations of evolutionary distance (which is way too complex and too little understood to compute en masse)}
%
\fi
%
between genomes}.
%
%\haruncor{We
%provide software that seamlessly integrates
%GenomeAtScale with other elements of a general genomics project flow, such as
%reading the input genome data in established formats.}
%
By maintaining compatibility
with standard bioinformatics data formats, we allow for GenomeAtScale to be
seamlessly integrated into existing analysis pipelines. Our performance analysis
illustrates the scalability of our solutions. In addition, we 
benchmark
GenomeAtScale on both large (2,580 human RNASeq experiments)
and massive (almost all public bacterial and viral whole-genome sequencing experiments) scales, and we make this data publicly available to foster
high-performance distributed genomics research.
%
%To the best of our knowledge, \emph{our work is the first to consider a
%distributed exact computation of the Jaccard similarity for genome distance
%derivations}. 
%
Despite our focus on genomics data, the algebraic formulation and
the implementation of our routines for deriving Jaccard measures are generic
and can be directly used in other settings. 

Specifically, 
our work makes the following contributions.
\begin{itemize}[leftmargin=0.5em]
\item We design \textbf{SimilarityAtScale}, the first communication-efficient
distributed algorithm to compute the Jaccard similarity index and distance.
\item We use our algorithm as a backend to \textbf{GenomeAtScale}, the first tool
that enables fast, scalable, accurate, and large-scale derivations of Jaccard distances
between high-throughput whole-genome sequencing samples. 
\item We ensure that SimilarityAtScale is generic and can be reused in any
other related problem. 
We
overview the
relevance of Jaccard measures in
data mining
applications.
\item We support our algorithm with a theoretical
analysis of the communication costs
and parallel scaling efficiency.
\item We evaluate GenomeAtScale on real genomic datasets,
showing that it enables large-scale 
genomic analysis. We scale our runs to up to 1024 compute
nodes, which is the largest scale that we know of 
for genetic distance computations~\cite{ondov2016mash,choi2018libra}.
\end{itemize}

\emph{The whole implementation of GenomeAtScale, as
well as the analysis outcomes for established real genome datasets, are
publicly available in order to enable interpretability and reproducibility, and
to facilitate 
its integration into current and future genomics analysis pipelines.}

% \section{Why Does Jaccard similarity Matter?}

\section{Jaccard Measures: Definitions and Importance}
\label{sec:defs_importance}

We start with defining the Jaccard measures and with discussing their
applications in different domains.
The most important symbols used in this work are listed in Table~\ref{tab:symbols}.
While we focus on high-performance
computations of distances between genome sequences, our design and
implementation are generic and applicable to any other use case.

\begin{table}[h]
\centering
%\footnotesize
\scriptsize
%\ssmall
\sf
%
%\begin{tabular}{@{}l|ll@{}}
%
\begin{tabular}{@{}ll@{}}
\toprule
\multicolumn{2}{c}{\textbf{General Jaccard measures:}} \\
\midrule
$J(X,Y)$ & The Jaccard similarity index of sets $X$ and $Y$. \\
$d_J(X,Y)$ & The Jaccard distance between sets $X$ and $Y$; $d_J = 1 - J$. \\
%
%\multirow{3}{*}{\begin{turn}{90}\shortstack{{Graph}\\{struct.}}\end{turn}}
%
%\multirow{5}{*}{\begin{turn}{90}\shortstack{{Algebraic}\\{notation}}\end{turn}}
%
\midrule
\multicolumn{2}{c}{\textbf{Algebraic Jaccard measures (\emph{SimilarityAtScale}), details in~\cref{sec:formulation}:}} \\
\midrule
$m,n$ & \makecell[l]{The number of possible data values (attributes) and data samples, \\one sample contains zero, one, or more values (attributes).}\\
$\otimes, \odot$ & \makecell[l]{Matrix-vector (MV) and vector dot products.} \\
$\mat{A}$ & \makecell[l]{The \emph{indicator matrix} (it determines the presence of data values\\in data samples), $\mat{A} \in \mathbb{B}^{m \times n}, \mathbb{B} = \{0, 1\}$.} \\
$\mat{S}, \mat{D}$ & \makecell[l]{The \emph{similarity} and \emph{distance} matrices with the values of Jaccard\\measures between all pairs of data samples; $\mat{S}, \mat{D} \in \mathbb{R}^{n \times n}$.} \\
$\mat{B}, \mat{C}$ & Intermediate matrices used
to compute $\mat{S}$, $\mat{B}, \mat{C} \in \mathbb{N}^{n \times n}$. \\
%$b$ & Number of batches.\\
%                    & $\odot$ &Elementwise vector-vector (VV) multiplication.\\
%                    & $\mat{A},\mat{A'}$& $G$'s adjacency matrix and the transformed $\mat{A}$ used in~\cref{sec:tropical}.\\
%                    & $\mat{x}_k,\mat{f}_k$& The result of $\mat{A} \otimes \mat{f}_k$; the frontier of vertices in iteration~$k$.\\
%                    & $\mat{g}_k,\mat{p}_k$& Helper vectors used in boolean, real, and sel-max semirings; see~\cref{sec:accelerate_bfs_variants}.\\
%                    & $\mat{d}, \mat{p}$ & Distances (to the root) and parents (in th BFS traversal tree). \\
%                    & $DP(\mat{d})$ & A transformation that derives vertex parents $\mat{p}$ from $\mat{d}$. \\
%                    & $\mat{x}^s,\mat{x}_k$ &$s$th element of a vector $\mat{x}$ and a vector $\mat{x}$ in $k$th iteration, respectively.\\
%                    & $\mat{x}_i$ &Vector x in $i$th iteration of computation.\\
%                    & $\overline{\mat{x}}$ &Element-wise negation of a vector $\mat{x}$.\\
%                   \midrule
%\multirow{3}{*}{\begin{turn}{90}\shortstack{Various}\end{turn}}
%                    & $T,W$ &The number of threads; work complexity of a given scheme.\\
%                    & $C,n_c$ &Chunk height; the number of chunks in SlimSell or Sell-$C$-$\sigma$.\\
%                    & $\sigma$ &Sorting scope in SlimSell and Sell-$C$-$\sigma$ ($\sigma \in [1,n]$).\\
%
\midrule
\multicolumn{2}{c}{\textbf{Genomics and metagenomics computations (\emph{GenomeAtScale}):}} \\
\midrule
$k$ & The number of single nucleotides in a genome subsequence.\\
$m,n$ & The number of analyzed $k$-mers and genome data samples.\\
%
%\midrule
%
%\multicolumn{2}{c}{\textbf{Design, implementation, and evaluation:}} \\
%
%\midrule
%
%$P$ & Number of processes.\\
%
%\midrule
%
\bottomrule
\end{tabular}
\vspace{-0.5em}
\caption{The most important symbols used in the paper.}
\label{tab:symbols}
\vspace{-1.5em}
\end{table}

\subsection{Fundamental Definitions}
\label{sec:basic_defs}

The \emph{Jaccard similarity index}~\cite{levandowsky1971distance} is a
statistic used to assess the similarity of sets. It is defined as the ratio of
the set intersection cardinality and the set union cardinality,

\begin{align*}
J(X,Y) = \frac{|X \cap Y|}{|X \cup Y|} = \frac{|X \cap Y|}{|X| + |Y| - |X \cap Y|},
\end{align*}
where $X$ and $Y$ are sample sets. If both $X$ and $Y$ are empty, the index is
defined as $J(X,Y) = 1$. One can then use $J(X,Y)$ to define the \emph{Jaccard
distance} $d_J(X,Y) = 1 - J(X,Y)$, which assesses the \emph{dissimilarity} of
sets and which is a proper \emph{metric} (on the collection of all finite
sets).

We are interested in the computation of similarity between all pairs of a collection of sets (in the context of genomics, this collection contains samples from one or multiple genomes).
Let $\mathcal{X}=\{X_1,\ldots, X_n\}$ be the set of \emph{data samples}.
Each data sample consists of a set of positive integers up to $m$, so $X_i \subseteq \{1,\ldots, m\}$.
We
seek to compute the Jaccard similarity for each pair of data samples:
\begin{align}
J(X_i, X_j) = \frac{|X_i \cap X_j|}{|X_i \cup X_j|},\quad i,j \in \{1, \ldots , n\}. \label{eq:pairwise-j}
\end{align}

\subsection{Computing Genetic Distances}
\label{sec:fundamentals_genomics}
Due to gaps in knowledge and the general complexity of computing evolutionary
distances between samples of genetic sequences, much effort has
been dedicated to efficiently computing 
accurate proxies for genetic distance~\cite{zielezinski2017alignment}.
The majority of these
works fall into one of two categories: \emph{alignment-based} and
\emph{alignment-free} methods. When comparing two (or more) sequences,
alignment-based methods 
explicitly take into account the ordering of the characters when determining the
similarity. 
More specifically, such methods assume a certain mutation/evolutionary model between the sequences and compute
a minimal-cost sequence of edits until all compared sequences converge.
%
%assume that the compared sequences \emph{are both
%ordered}, and this order is explicitly taken into account while determining the
%similarity.
%
This family of methods forms a mature area of research. An
established alignment-based tool for deriving genome distances is
BLAST~\cite{altschul1990basic}.
While highly accurate, these methods are computationally intractable when comparing
sets of high-throughput sequencing data.
Contrarily, alignment-free methods do not consider the order of individual
bases or amino acids while computing the distances between analyzed sequences. These
methods have been recently proposed and are in general much faster than
alignment-based designs~\cite{zielezinski2017alignment}. Both exact and
approximate variants were investigated, including the mapping of $k$-mers to
common ancestors on a taxonomic tree~\cite{wood2014kraken}, and the approximation of the
Jaccard similarity index using 
minwise hashing~\cite{ondov2016mash}.  
To the best of our knowledge, there is no
previous work that enables distributed computation of a genetic distance that is fast,
accurate, and \emph{scales to 
massive sets of sequencing samples}.

\iffalse % REVISION

\revision{A recent closely related work, BELLA~\cite{bella2019} leverages multiplication of sparse matrices for the \haruncor{alignment problem in}{computation of pairwise overlaps between long reads as a part of} genome assembly via Overlap Layout Consensus (OLC)~\cite{chu2016innovations}\haruncor{
BELLA calculates the alignment between long reads,}{, }and has been extended to support parallel execution~\cite{ellis2019dibella}.
Our work\harun{, on the other hand} targets whole genome comparison, \haruncor{processing much larger datasets than long reads}{computing distances between much larger, entire read sets}.
Consequently, our parallelization strategy is very different.
More details comparing these two approaches are in Section~\ref{sec:discussion}.}
%
%focuses on the problem of overlap detection in Overlap-Layout-Consensus (OLC) paradigm. This work is closely related to ours, and we discuss the fundamental differences between the two in Section \ref{sec:discussion}.  
%
\revision{To the best of our knowledge}, there is no
previous work that enables distributed computation of a genetic distance that is fast,
accurate, and \emph{scales to 
massive sets of sequencing samples}.

\fi

\begin{table*}[t]
\centering
%\footnotesize
\scriptsize
%\ssmall
\sf
\begin{tabular}{@{}llllll@{}}
\toprule
\textbf{Tool} & \textbf{\# compute nodes} & \textbf{\# samples} & \textbf{Raw input data size} & \textbf{Preprocessed data size} & \textbf{Similarity} \\
\midrule
DSM~\cite{seth2014exploration} & $1$        & $435$     & $3.3$TB           & N/A$^\ddag$    & Jaccard \\
Mash~\cite{ondov2016mash}      & $1$        & $54,118$  & N/A$^\dagger$     & $674$GB        & Jaccard (MinHash) \\
Libra~\cite{choi2018libra}     & $10$       & $40$      & $372$GB           & N/A$^\ddag$    & Cosine \\ \hline
GenomeAtScale                  & $1024$     & $446,506$ & $170$TB           & $1.8$TB        & Jaccard \\
\midrule
\bottomrule
\end{tabular}
\vspace{-0.5em}
\caption{\textbf{
%\edgar{Does it make sense to include BELLA in above table?} \harun{Harun: I don't think so. The samples in BELLA are a different type from the other studies, so the comparison would be misleading.}
Comparison of scales of different alignment-free tools for deriving genetic distances.}
Raw input data refers to high-throughput sequencing data, while preprocessed data
refers to cleaned and assembled long sequences. All sizes given refer to uncompressed FASTA files.
$\dagger$ Mash is constructed from assembled and curated reference genomes, where in some cases
the corresponding raw sequencing data files may not be available.
$\ddag$ DSM and Libra directly query raw sequencing data with no assembly step.
GenomeAtScale was computed from cleaned and assembled sequences (see Section~\ref{ref:datasets} for details).
%
%\haruncomment{PROCESSED SIZE FOR GENOMEATSCALE IS APPROXIMATED AS 2*COMPRESSED SIZE. COMPUTING EXACT UNCOMPRESSED SIZE NOW}
%
%$^*$This number is inferred from the provided description of the experimental configuration~\cite{ondov2016mash} and comparison to the used Biowulf cluster.
\textbf{GenomeAtScale 
is shown to achieve larges problem size and parallelism scales than
past approaches.}}
\label{tab:scalability_comparison}
%\vspace{-0.5em}
\end{table*}

A \emph{genome} is a collection of sequences defined on the alphabet of the \emph{nucleotides} adenine (A),
thymine (T), guanine (G), and cytosine (C) (each individual sequence is referred to as a \emph{chromosome}).
A $k$-mer is a subsequence of
length~$k$ of a given sequence. For example, in a sequence AATGTC, there are
four 3-mers (AAT, ATG, TGT, GTC) and three 4-mers (AATG, ATGT, TGTC).
%
%, and two
%5-mers (AATGT, ATGTC).
%
%\edgar{How does a sample below relate to a "long-read" above?}
%\harun{Harun: this is addressed in discussion.tex}
A common task in genomics is to reconstruct the \emph{assembly} of
chromosomes from one or more
species given \emph{high-throughput sequencing} samples.
These assembled sequences allow for accurate, alignment-based methods to be used
when assessing the similarity between samples. Yet, the high computational
cost of assembly has meant that the vast majority of available sequencing data
has remained unassembled~\cite{stephens2015biggenomical}. Thus, there has been a growing
interest in methods that enable such comparisons to be made on representations
of sequencing data without requiring prior assembly.
Alignment-free
methods typically represent a sequencing sample~$i$ as a set~$X_i$ of
its respective $k$-mers. From this, one may compute the genetic distance to another
sample~$X_j$ via the Jaccard similarity of $X_i$ and $X_j$~\cite{ondov2016mash},
or map each $k$-mer in $X_i$ to a database of labels for classification~\cite{wood2014kraken}.
The distance matrix $1 - J(i,j)$ may then be used for subsequent downstream tasks,
including the clustering of samples for the construction of phylogenetic trees~\cite{saitou1987neighbor}, to aid the selection
of related samples for joint metagenomic analysis~\cite{popic2018fast}, or to aid
the construction of guide trees for large-scale multiple sequence alignment~\cite{armstrong2019progressive} (see Figure~\ref{fig:flow}).

We briefly discuss the scalability of different alignment-free tools
for deriving genetic distances, and compare them to the proposed GenomeAtScale. The considered aspects are (1) the
  size of processed genome data (both the input size and the number of samples in this input), (2) the amount of used compute resources,
  and (3) the used measure of similarity and its accuracy. A comparison is presented
  in Table~\ref{tab:scalability_comparison}.
GenomeAtScale delivers the largest scale of genome distance computations in the all considered aspects.
We will describe the evaluation in detail in Section~\ref{sec:eval}.

\subsection{General Data Science: Clustering}

As the Jaccard distance~$d_J$ is formally a metric, it can be used with many
clustering routines. For example, one can use it with centroid-based clustering
such as the popular $k$-means algorithm when deriving distances between data
points and centroids of clusters~\cite{ferdous2009efficient, wu2014service}.
It can also be used with hierarchical
clustering and other methods~\cite{theodoridis2009pattern, dong2006hierarchical}.
The advantage of using $d_J$ is that it is straightforwardly applicable to
categorical data that does not consist of numbers but rather attributes that
may be present or absent~\cite{morzy2001scalable, schaeffer2007graph}.

\subsection{General Data Science: Anomaly Detection}

Another use case for the Jaccard distance is anomaly detection through
proximity-based outlier detection~\cite{kotu2018data}. This application is
particularly useful when the analyzed data contains binary or categorical
values.

\subsection{Machine Learning: Object Detection}

In object detection, the Jaccard similarity is referred to as
\emph{Intersection over Union} and it is described as the most popular
evaluation metric~\cite{rezatofighi2019generalized}. Here, sets $X$ and $Y$
model two bounding boxes: a ground-truth bounding box around an object
to be localized in a picture and a predicted bounding box. $|X
\cap Y|$ is the overlap area of these two boxes; $|X \cup Y|$ constitutes
the union area. The ratio of these two values assesses how well a predicted
box matches the ideal box.

\subsection{Graph Analytics and Graph Mining}

The Jaccard similarity is also used in graph analytics and
mining~\cite{besta2019substream, besta2019graph, gianinazzi2018communication,
solomonik2017scaling, besta2017push, besta2017slimsell, besta2015accelerating,
besta2015active, besta2018log, besta2019demystifying, besta2018survey,
besta2019slim}, to compute the similarity of any two vertices $v,u$ using only
the graph adjacency information. This similarity can be defined as $|N(v) \cap
N(u)| / |N(v) \cup N(u)|$, where $N(v)$ and $N(u)$ are sets with neighboring
vertices of $v$ and $u$, respectively~\cite{schaeffer2007graph}. Vertex
similarity is often used as a building block of more complex algorithms. One
example is Jarvis-Patrick graph clustering~\cite{jarvis1973clustering}, where
the similarity value determines whether $v$ and $u$ are in the same cluster.
Other examples include discovering missing links~\cite{chen2012discovering} or
predicting which links will appear in dynamic graphs~\cite{zhou2019attacking, besta2019practice}.

\subsection{Information Retrieval}

In the information retrieval context, $J(X,Y)$ can be defined as the ratio of
the counts of common and unique words in sets $X$ and $Y$ that model two
documents.  Here, $J(X,Y)$ assesses the similarity of two such documents.  For
example, text2vec, an R package for text analysis and natural language
processing~\cite{selivanov2016text2vec}, uses the Jaccard distance for this
purpose.

\section{Communication-Efficient Jaccard Measures}

We now describe \textbf{SimilarityAtScale}, a distributed
algorithm for deriving Jaccard measures
based on sparse linear algebra.
In Section~\ref{sec:genomeatscale}, we 
apply our algorithm to genomics.

\subsection{Algebraic Formulation of Jaccard Measures}
\label{sec:formulation}

We first provide an algebraic formulation of the Jaccard measures
(the following description uses definitions from~\cref{sec:basic_defs}).
We define an \emph{indicator matrix}~$\mat{A} \in
\mathbb{B}^{m \times n}$, where $\mathbb{B} = \{0,1\}$. We have
\vspace{-1em}
\[{a}_{ij} = \begin{cases} 1 & : i \in X_j \\ 0 & : \text{otherwise} \end{cases}.\]
The matrix~$\mat{A}$ determines the presence of data values in data samples.
We seek to obtain the \emph{similarity matrix},
$\mat{S}\in \mathbb{R}^{n\times n}$ defined to obtain the similarities described in~\eqref{eq:pairwise-j},
\[s_{ij} = J(X_i, X_j) = \frac{|X_i \cap X_j|}{|X_i \cup X_j|}.\]
To compute $\mat{S}$, it suffices to form matrices $\mat{B},\mat{C}\in \mathbb{N}^{n\times n}$,
which describe the cardinalities of the intersections and unions of each pair of data samples, respectively.
The computation of $\mat{B}$ is most critical, and can be described as a sparse matrix--matrix product,
\[b_{ij} = |X_i \cap X_j| = \sum_k a_{ki}a_{kj}, \text{ so } \mat{B} = \mat{A}^T \mat{A}.\]
The matrix $\mat{C}$ can subsequently be obtained via $\mat{B}$
(we use $\vcr{\hat{a}} = (\hat{a}_1, ..., \hat{a}_n)^T$ to simplify notation):

\begin{align*}
c_{ij} &= |X_i \cup X_j| = |X_i| + |X_j| - |X_i \cap X_j| \\
&= \hat{a}_i + \hat{a}_j- b_{ij}, 
 \text{ where } \hat{a}_i = \sum_{k} a_{ki}.
\end{align*}

\noindent
The
similarity and the distance matrices~$\mat{S}$,
$\mat{D}$ are 
given by
\begin{gather}
s_{ij} = b_{ij} / c_{ij},
%\quad i \in \{1, ..., n\}, j \in \{1, ..., n\}. \\
d_{ij} = 1 - s_{ij},\quad i,j \in \{1, ..., n\},\label{eq:S_def}
% j \in \{1, ..., n\}.
%
\end{gather}
The formulation and the algorithm are generic and can be used in any setting
where compared data samples are \emph{categorical} (i.e., they consist of
\emph{attributes} that may or may not be present in a given sample). For
example, a given genome data sample usually contains some number of $k$-mers
that form a subset of all possible $k$-mers.  Still, most numerical data can be
transformed into the categorical form. For example, the
neighborhood~$N(v)$ of a given graph vertex~$v$ usually contains integer vertex IDs
($N(v) \subset \mathbb{N}$). 
Hence,
one can model all
neighborhoods with the adjacency matrix~\cite{besta2017slimsell}.

\subsection{Algorithm Description}

The formulation of Jaccard similarity computation via sparse linear algebra derived in
Section~\cref{sec:formulation} is succinct, but retains some algorithmic challenges.
First, the data samples (nonzeros contained in indicator matrix $\mat{A}$) may not 
simultaneously fit in the combined memory of all nodes on a distributed-memory system.
Second, the similarity matrix $\mat{S}$ may not fit in the memory of a single node,
but should generally fit in the combined memory of the parallel system.
\emph{Further, the most significant computational challenge is that the 
indicator matrix $\mat{A}$ is incredibly sparse for genomic similarity problems.}
The range of $k$-mers generally extends to $m=4^{30}$.
This means that $\mat{A}$ is very hypersparse~\cite{buluc2008representation}, i.e., the overwhelming majority
of its rows are entirely zero.
To resolve this challenge, the SimilarityAtScale algorithm makes use of three techniques:
\begin{enumerate}[leftmargin=1em]
\item subdividing $\mat{A}$'s rows into batches, each with $\tilde{m}$ rows,
\item filtering zero rows within each batch using a distributed sparse vector,
\item masking row segments into bit vectors.
\end{enumerate}
We now describe in detail how these techniques are deployed.  \hl{A high-level
pseudocode can be found in Listing~}\colorbox{white}{\ref{lst:high-alg}}
while more details are provided in Listing~\ref{lst:alg}.

\ifrev
\marginpar{\vspace{-3em}\colorbox{orange}{\textbf{ALL}}}
\fi

\begin{lstlisting}[aboveskip=1em, belowskip=-2.5em, float=!h,label=lst:high-alg,caption=\hl{\textbf{High-level pseudocode} of the SimilarityAtScale algorithm.
All mathematical symbols are defined in Section~}\colorbox{white}{\ref{sec:formulation}}\hl{ and listed in Table~}\colorbox{white}{\ref{tab:symbols}.}]
//For any details of specific structures or operations, see Listing |\ref{lst:alg}|
//Below, we refer to respective equations in the text

for each batch of the input matrix $\mat{A}$ { //Batches are defined in Eq.(|\ref{eq:batches_def}|).
  Read the next $l$-th batch $\mat{A}^{(l)}$ of $\mat{A}$
  Remove zero rows from $\mat{A}^{(l)}$ using a filter $\mat{f}^{(l)}$, the result is $\mat{\bar{A}}^{(l)}$
  //The filter $\mat{f}^{(l)}$ and the matrix $\mat{\bar{A}}^{(l)}$ are defined in Eq.(|\ref{eq:filter_def}|) and (|\ref{eq:A_bar_def}|).
  Compress $\mat{\bar{A}}^{(l)}$ with bitmasking, the result is $\mat{\hat{A}}^{(l)}$
  //The matrix $\mat{\hat{A}}^{(l)}$ and bitmasking are defined between Eq.(|\ref{eq:filter_def}|) and (|\ref{eq:A_bar_def}|).
  Compute the partial scores $\mat{S}^{(l)} = \mat{A}^{(l) T} \mat{A}^{(l)}$ and $\mat{\hat{a}}^{(l)}$
  //The partial scores are defined implicitly in Eq.(|\ref{eq:batches_int}|) and (|\ref{eq:final}|).
  Accumulate the partial scores into intermediate matrices $\mat{B}$ and $\mat{\hat{a}}$
  //Intermediate matrices $\mat{B}$ and $\mat{\hat{a}}$ are defined in Eq.(|\ref{eq:batches_int}|).
  //Recall that $\mat{B}$ describes the cardinalities of intersections of
  //pairs of data samples, and - together with $\mat{\hat{a}}$ - can be used
  //to describe the cardinalities of unions of pairs of data samples.
Derive the final similarity scores $\mat{S}$ based on $\mat{B}$ and $\mat{\hat{a}}$
//Batches $\mat{S}^{(l)}$ of $\mat{S}$ are defined in Eq.(|\ref{eq:final}|), $\mat{S}$ is defined in Eq.(|\ref{eq:S_def}|).
\end{lstlisting}

\begin{lstlisting}[aboveskip=1em, belowskip=-2.5em, float=!h,label=lst:alg,caption=The details of the SimilarityAtScale algorithm.
All mathematical symbols are defined in Section~\ref{sec:formulation} and listed in Table~\ref{tab:symbols}.]
|\vspace{0.25em}\tikzmarkin[set fill color=grayblack, set border color=white, above offset=0.25, below offset=-0.1]{bl1} \textcolor{white}{/*\quad\quad\quad\quad\quad\quad\  Notation remarks, description of input and output \ \quad\quad\quad\quad\quad\quad */} \tikzmarkend{bl1}|
/* For simplicity, we use the same symbols for the variables that
 * correspond to the introduced mathematical objects: matrices |\textbf{A}|, |\textbf{B}|,
 * |\textbf{C}|, |\textbf{S}|; a number of all data samples $n$ and attributes $m$. 
 * |\textbf{Input:}| ``|\ul{files}|'': an array of pointers to $n$ files, where one file
 * contains data values from one data sample $X_i\ (i \in \{1, ..., n\})$.
 * ``|\ul{batch\_cnt}|'': the number of batches into which we partition
 * the derivation and processing of matrices |\textbf{A}|, |\textbf{B}|, and |\textbf{C}|.
 * ``|\ul{max\_val}|'': the maximum value across all samples $X_i$.
 * ``|\ul{bitmask}|'': a bitmask used to compress the (|boolean|) input data
 * and reduce the memory footprint in |\textbf{A}|. ``|\ul{comm}|'' is an object with
 * details of the distributed execution (e.g., process count). 
 * ``|\ul{Value}|'' is an opaque object that represents an arbitrary value
 * possibly contained in input data (e.g., a number, a letter).
 * |\textbf{Output:}| The Jaccard similarity matrix |\textbf{S}| (it can be trivially
 * transformed into the Jaccard distance matrix |\textbf{D}|). */

|\vspace{0.25em}\tikzmarkin[set fill color=grayblack, set border color=white, above offset=0.25, below offset=-0.1]{bl2} \textcolor{white}{/*\quad A part that combines all elements used to derive the Jaccard measures \quad */} \tikzmarkend{bl2}|
typedef vector<pair<index, data>> Vector;

// Derive and return the Jaccard similarity matrix.
Matrix* SimilarityAtScale(File* files, int $n$, int $m$,
    int bitmask, Comm* comm, int batch_cnt) {
  int batch_cnt_tot = batch_cnt + ($m$ % batch_cnt) > 0;
  //"DMatrix"/"DVector" indicate a distributed matrix/vector
  DMatrix |\textbf{A}|, |\textbf{B}|, |\textbf{C}|; // Declare opaque matrix objects.
  DVector |\textbf{f}|; // Declare a temporary data structure for input.

  for(int i = 0; i < batch_cnt_tot; i++) {
    readFiles(files, $n$, comm, &f); // Read input in batches.
    // Compress each input batch and remove zero rows.
    preprocessInput($n$, $m$, batch_cnt, bitmask, &|\textbf{f}|, &|\textbf{A}|); 
    jaccardAccumulate(&|\textbf{A}|, &|\textbf{B}|, &|\textbf{C}|); // Construct matrices |\textbf{A}|, |\textbf{B}|, |\textbf{C}|.
  }
|\label{ln:man_1_1}|  |\textbf{C}|["ij"] -= |\textbf{B}|["ij"];
|\label{ln:man_1_2}|  |\textbf{S}|["ij"] += |\textbf{B}|["ij"] / |\textbf{C}|["ij"]; // Derive the similarity matrix.
  return &|\textbf{S}|; // Return the Jaccard similarity matrix.
}

|\vspace{0.25em}\tikzmarkin[set fill color=grayblack, set border color=white, above offset=0.25, below offset=-0.1]{bl3} \textcolor{white}{/* \quad\quad\quad\quad\quad\quad\quad\quad\quad\quad\quad\quad\quad\quad Loading input files \quad\quad\quad\quad\quad\quad\quad\quad\quad\quad\quad\quad\quad */} \tikzmarkend{bl3}|
void readFiles(File* files, int $n$, Comm* comm, DVector* f) { 
  // This function is executed by each process in parallel.
  Vector data_sample;
  for(int i = comm->my_rank; i < $n$; i += comm->num_procs) {
    // One file line contains one data value.
    Value val = files[i]->read_file_line();
    // Store the value in a tuple; <index=val,data=1>.
    data_sample.push(val, 1); 
  }
  // Bulk update a structure with loaded data (|\textbf{f}|[data_value.index]=1).
  // We use a write function on the vector in which
  // all processes update the loaded data in parallel.
  |\textbf{f}|.write(data_sample);
}

|\vspace{0.25em}\tikzmarkin[set fill color=grayblack, set border color=white, above offset=0.25, below offset=-0.1]{bl4} \textcolor{white}{/* \quad\quad\quad Preprocessing (removing zero rows, compression using the bitmask) \quad\quad */} \tikzmarkend{bl4}|
void preprocessInput(int $n$, int $m$, int batch_cnt, int bitmask,
                       Vector* R, Matrix* |\textbf{A}|) {
  int len_bm = length(bitmask); // Get the bitmask length.
  // This function is executed by each process in parallel.
  // Get the non-zero data indices.
|\label{alg:get_nonzeros}|  Vector* nonzero_data = get_all_nonzero_pairs(R); 
  // ``nonzero_data'' effectively represents the non-zero rows. 
  for(int i = comm->my_rank; i < $n$; i += comm->num_procs) {
    int j = 0; int mask = 0;
    Vector masked_data_sample;
    while (j < $m$) {
      Value val = data_sample[j++].first;
      int l = 0;
      while (j < nonzero_data.size && l++ < len_bm) {
        if (nonzero_data.index == (val % ($m$ / batch_cnt))) {
          // The iteration follows the column-major order
          // (this reflects our implementation).
          mask $\vert$= ((bitmask)1) << ((j % ($m$ / batch_cnt))) % len_bm);
      } }
      if (mask) masked_data_sample.push(mask_index, mask);
  } } 
  write(|\textbf{A}|, masked_data_sample); // Bulk update |\textbf{A}|.
}

|\vspace{0.25em}\tikzmarkin[set fill color=grayblack, set border color=white, above offset=0.25, below offset=-0.1]{bl5} \textcolor{white}{/* \quad\quad\quad\quad\quad\quad Deriving intermediate matrices \textbf{B} and \textbf{C} in batches \quad\quad\quad\quad\quad */} \tikzmarkend{bl5}|
void jaccardAccumulate(Matrix* |\textbf{A}|, Matrix* |\textbf{B}|, Matrix* |\textbf{C}|) {
  // ``popcount'' counts the number of ``ones'' in a given row/column.
  // The operations below follow the specification in |\cref{sec:formulation}|.
|\label{ln:man_2_1}|  |\textbf{B}|["ij"] = popcount(|\textbf{A}|["ki"] & |\textbf{A}|["kj"]);
  |\textbf{v}|["i"] += popcount(|\textbf{A}|["ki"]);
|\label{ln:man_2_2}|  |\textbf{C}|["ij"] += |\textbf{v}|["i"] + |\textbf{v}|["j"];
}
\end{lstlisting}

\ifrev
\marginpar{\vspace{-5em}\colorbox{orange}{\textbf{ALL}}}
\fi

We subdivide the indicator into batches $r=m/\tilde{m}$ (to simplify the presentation we assume $m$ divides into $\tilde{m}$) batches:
\begin{gather}\mat{A} = \begin{bmatrix} \mat{A}^{(1)} \\ \vdots \\\mat{A}^{(r)} \end{bmatrix}, \text{ where }\mat{A}^{(l)}\in \mathbb{B}^{\tilde{m}\times n}, \forall l\in\{1,\ldots, r\}.\label{eq:batches_def}\end{gather}
To obtain the similarity matrix~$\mathbf{S}$, we need to obtain $\vcr{\hat{a}}$ and $\mat{B}$,
which can be combined by accumulation of contributions from each batch,
\small
\begin{gather}\mat{B} = \sum_{l=1}^{r} \mat{A}^{(l)}{}^T \mat{A}^{(l)}, \quad \vcr{\hat{a}} = \sum_{l=1}^{r}\vcr{\hat{a}}^{(l)}, \text{ where } \hat{a}^{(l)}_i = \sum_k a^{(l)}_{ki}.\label{eq:batches_int}\end{gather}\normalsize

To filter out nonzero rows in a batch, we construct a sparse vector $\vcr{f}^{(l)}\in\mathbb{B}^{\tilde{m}}$
that acts as a filter,
\begin{gather} f_k^{(l)} = \begin{cases} 1 & : \exists i\in\{1,\ldots, n\}, a^{(l)}_{ki} \neq 0 \\ 0 & : \text{otherwise} \end{cases}.\label{eq:filter_def}\end{gather}
%The computation of $\vcr{f}^{(l)}$ is done in a parallel manner via sparse data input primitives provided in the Cyclops library, which are described in more detail in Section~\cref{subsec:impl}.
Given the prefix sum $\vcr{p}^{(l)}$ of $\vcr{f}^{(l)}$, the batch of the indicator matrix $\mat{A}^{(l)}$ can be reduced to a matrix $\mat{\bar{A}}^{(l)}$ that contains only nonzero rows,
\begin{gather}\bar{a}^{(l)}_{p^{(l)}_ki} = a^{(l)}_{ki}.\label{eq:A_bar_def}\end{gather}
Subsequently, it suffices to work with $\mat{\bar{A}}^{(l)}$ since \[\mat{A}^{(l)}{}^T \mat{A}^{(l)} = \mat{\bar{A}}^{(l)}{}^T \mat{\bar{A}}^{(l)}.\]

Even after removal of nonzero rows in each batch of the indicator matrix, it helps to further reduce the number of rows in each $\mat{\bar{A}}^{(l)}$.
The meta-data in both COO and CSR formats necessary to store each nonzero corresponds to a 32-bit or 64-bit integer.
In the CSR layout, the same amount of meta-data is necessary to store each ``row start'' count.
We reduce the latter overhead by leveraging the fact that each binary value only requires one bit of data.
In particular, we encode segments of $b$ elements of each column of $\mat{\bar{A}}^{(l)}$ in a $b$-bit bitmask.
A natural choice is $b=32$ or $b=64$, which increases the storage necessary for each nonzero by no more than $2{-}3\times$, while reducing the number of rows (and consequently row-start counts in the CSR representation) by $b$, as well as potentially reducing the number of actual nonzeros stored.
The resulting matrix $\mat{\hat{A}}^{(l)}\in\mathbb{S}^{(\tilde{m}/b)\times n}$ where $\mathbb{S}=\{0,\ldots, 2^{b}-1\}$, can be used effectively to compute the similarity matrix.
In particular, for $\mat{S}^{(l)} = \mat{A}^{(l)}{}^T \mat{A}^{(l)}$, we have
\begin{gather}s^{(l)}_{ij} = \sum_k \text{popcount}(\hat{a}^{(l)}_{ki} \wedge \hat{a}^{(l)}_{kj}),\label{eq:final}\end{gather}
where $\text{popcount}(x)$ counts the number of set bits in $x$.

\subsection{Parallelization and Analysis}
\label{sec:par_analysis}

We assume without loss of generality that the rows and columns of $\mat{A}$ are
randomly ordered, which can be enforced via a random reordering.
Thus, the cost of computing each batch is roughly the
same.
Let $h=\tilde{m}/b$ so that $\mat{R}=\mat{\hat{A}}^{(l)}\in\mathbb{S}^{h\times n}$, and let $z$ be the number of nonzeros in $\mat{R}$.
We assume $b=O(1)$ so that the cost of popcount is $O(1)$ arithmetic operations.
We first analyze the cost of the sparse matrix--matrix product, then quantify the overheads of the initial compression steps.
We consider a $p$ processor Bulk Synchronous Parallel (BSP) model~\cite{valiant1990bridging,skillicorn1997questions}, where each processor can store $M$ words of data, the cost of a superstep (global synchronization) is $\alpha$, the per byte bandwidth cost is $\beta$, and each arithmetic operation costs $\gamma$.
We assume $\alpha \geq \beta \geq \gamma$.

Our parallelization and analysis of the sparse matrix--matrix product follows known communication-avoiding techniques for (sparse) matrix--matrix multiplication~\cite{SD_EUROPAR_2011,Solomonik:2017:SBC:3126908.3126971,azad2016exploiting,koanantakool2016communication,Ballard:2013:COP:2486159.2486196,gustavson1978two,DBLP:journals/corr/abs-1109-3739, kwasniewski2019red}.
Given enough memory to store $1\leq c \leq p$ copies of $\mat{B}$, i.e., $M=\Omega(cn^2/p)$, we define a $\sqrt{p/c}\times \sqrt{p/c} \times c$ processor grid.
On each $\sqrt{p/c}\times \sqrt{p/c}$ subgrid, we compute $1/c$th of the contributions to $\mat{B}$ from a given batch of the indicator matrix, $\mat{R}$.
Each processor then needs to compute $\mat{R}^{(s,t)}{}^T\mat{R}^{(s,v)}$, where each $\mat{R}^{(i,j)}$ is a $h/c \times n\sqrt{c/p}$ block of $\mat{R}$.
For a sufficiently large $z,c,p$, w.h.p., \#nonzeros in each $\mat{R}^{(i,j)}$ is $O(z/\sqrt{cp})$.
Using a SUMMA~\cite{gustavson1978two,DBLP:journals/corr/abs-1109-3739,Geijn:SUMMA:1997} algorithm, computing $\mat{R}^{(s,\star)}{}^T\mat{R}^{(s,\star)}$ on the $s$th layer of the processor grid takes $O(1+z/(M\sqrt{cp}))$ BSP supersteps.
Finally, assuming $c>1$, one needs a reduction to sum the contributions to $\mat{C}$ for each layer, which requires $O(1)$ supersteps, where each processor sends/receives at most $O(cn^2/p)$ data.
The overall BSP communication cost of our algorithm is
\[O\bigg( \Big(1+\frac{z}{M\sqrt{cp}}\Big)\cdot \alpha + \Big(\frac{z}{\sqrt{cp}} + \frac{cn^2}{p}\Big)\cdot \beta\bigg).\]
The number of arithmetic operations depends on the sparsity structure (i.e., if some rows of $\mat{R}$ are dense and others mostly zero more operations are required per nonzero than if the number of nonzeros per row is constant).
For a sufficiently large $z,c,p$ these total number of arithmetic operations, $F$, will be evenly distributed among processors, yielding a BSP arithmetic cost of $O((F/p)\cdot \gamma)$.

We allow each processor to read an independent set of data samples from disk for each batch.
Assuming $n\gg p$ and that no row contains more than $O(z/(p\log p))$ nonzeros (the average is $O(n/z)$), each processor reads $O(z/p)$ nonzeros.
This assumption may be violated if columns of $\mat{A}$ have highly variable density and $p$ approaches $n$, in which case the resulting load imbalance would restrict the batch size and create some overhead.
To filter out nonzero rows, we require the computation of $\vcr{f}^{(l)}$ and its prefix sum $\vcr{p}^{(l)}$.
We perform a transposition of the initial data, so that each process collects all data in $m/(rp)$ of the $m/r$ rows in the batch, and combines it locally to identify nonzero rows.
If the number of nonzeros $z$ and total rows $bh$ is sufficiently large, i.e., $z,bh\gg p$, each processor receives $O(z/p)$ entries and ends up with $O(bh/p)$ of then nonzero rows of $\vcr{f}^{(l)}$.
Subsequently, a prefix sum of the nonzero entries of $\vcr{f}^{(l)}$ can be done with BSP cost,
%\[O(\log_M(\min(hb,p)) \cdot \alpha + (hb/p+\min(M,p))\cdot \beta).\]
\[O(\alpha + p\cdot \beta),\]
assuming $p=O(M)$.
%\edgar{need to check the above}
The nonzero entries received by each processor can then be mapped to the appropriate row in the $bh\times n$ matrix $\mat{\bar{A}}^{(l)}$ and sent back to the originating processor.
At that point, each processor can compress the columns of $\mat{\bar{A}}^{(l)}$ by a factor of $b$ to produce the $h\times n$ matrix $\mat{R}$.

The overall BSP cost per batch assuming $b=O(1)$ is
\begin{align*}
T(z,n,M,c,p) &= O\bigg(\Big(1+\frac{z}{M\sqrt{cp}}\Big)\cdot \alpha \\
&+ \Big(\frac{z}{\sqrt{cp}} + \frac{cn^2}{p} + p\Big)\cdot \beta + \frac{F}{p}\gamma \bigg).
\end{align*}
Generally, we pick the batch size to use all available memory, so $z=\Theta(Mp)$, and replicate $\mat{B}$ in so far as possible, so $c=\Theta(\min(p,Mp/n^2))$.
Given this, and assuming $p=O(M)$ and $M\leq n^2$, which is the ``memory-bound regime'', which is critical in all of our experiments, the above cost simplifies to
\begin{align*}
\tilde{T}(n,M,p) &= O\bigg(\frac{n}{\sqrt{M}}\cdot \alpha + n\sqrt{M}\cdot \beta + \frac{F}{p}\gamma \bigg).
\end{align*}
For a problem with $m$ rows and $Z$ nonzeros overall, requiring $G$ arithmetic operations overall, maximizing the batch size gives the total cost,
\begin{align*}
\frac{Z}{Mp}\tilde{T}(n,M,p) &= O\bigg(\frac{nZ}{pM^{3/2}}\cdot \alpha + \frac{nZ}{\sqrt{M}p}\cdot \beta + \frac{G}{p}\gamma \bigg).
\end{align*}
These costs are comparable to the ideal cost achieved by parallel dense matrix--matrix multiplication~\cite{greygeneral2010} in the memory-dependent regime, where $Z=n^2$ and $G=n^3$.

Given a problem where the similarity matrix fits in memory with $p_0$ processors, i.e., $M=n^2/p_0$, we can consider strong scaling where batch size is increased along with the number of processors, until the entire problem fits in one batch.
The parallel efficiency is then given by the ratio of BSP cost for computing a batch with $z_0=O(n^2)$ nonzeros and $h_0$ nonzero rows with $p_0$ processors to computing a larger batch with $z=O(n^2\cdot p/p_0)$ nonzeros and $h=h_0\cdot p/p_0$ nonzero rows using up to $p=O(\min(M,n))$ processors,
\begin{align*}
E_p = &\frac{T(z_0,n,n^2/p_0,1,p_0)}{T(pz_0/p_0,n,n^2/p_0,p/p_0,p)}= O(1).
\end{align*}
Thus, our algorithm can theoretically achieve perfect strong scalability so long as the load balance assumptions are maintained.
These assumptions hold given either balanced density among data samples or a sufficiently large number of data samples, and so long as the number of processors does not exceed the local memory or the dimension $n$.
%parallel algorithm achieves perfect strong scaling up until the entire problem can be performed with a single batch, and so long as the load balance assumptions are maintained.

%\begin{gather}
%%
%c_{ij} = \text{nnz}(a_{i\cdot}^T \odot a_{j\cdot}),\quad i \in \{1, ..., n\}, j \in \{1, ..., \widetilde{m}\}
%%
%\end{gather}
%%
%TODO: need to define the same for B, after it is defined in the previous section?

\subsection{Algebraic Jaccard for Different Problems}

We briefly explain how to use the algebraic formulation of the Jaccard measures
in selected problems from Section~\ref{sec:defs_importance}. The key part is to
properly identify and encode data values and data samples within the indicator
matrix~$\mathbf{A}$. We illustrate this for selected problems in
Table~\ref{tab:framing}.

\begin{table}[h]
\centering
%\footnotesize
\scriptsize
%\ssmall
\sf
\begin{tabular}{@{}lll@{}}
\toprule
\textbf{Computational problem} & \textbf{One row of $\mathbf{A}$} & \textbf{One column of $\textbf{A}$} \\
\midrule
Distance of genomes & One $k$-mer & One genome data sample \\
Similarity of vertices & Neighbors of one vertex & Neighbors of one vertex \\
Similarity of documents & One word & One document \\
Similarity of clusters & One vertex & One cluster \\
\bottomrule
\end{tabular}
\vspace{-0.5em}
\caption{Framing of the SimilarityAtScale algorithm for different computational problems.
$\mathbf{A} \in \{0,1\}^{m \times n}$ is the indicator matrix that determines the presence of data values in
the compared data samples, detailed in Section~\ref{sec:formulation}.}
\label{tab:framing}
%\vspace{-0.5em}
\end{table}

\begin{figure*}[t]
\vspace{-1.5em}
\centering
\includegraphics[width=1.0\textwidth]{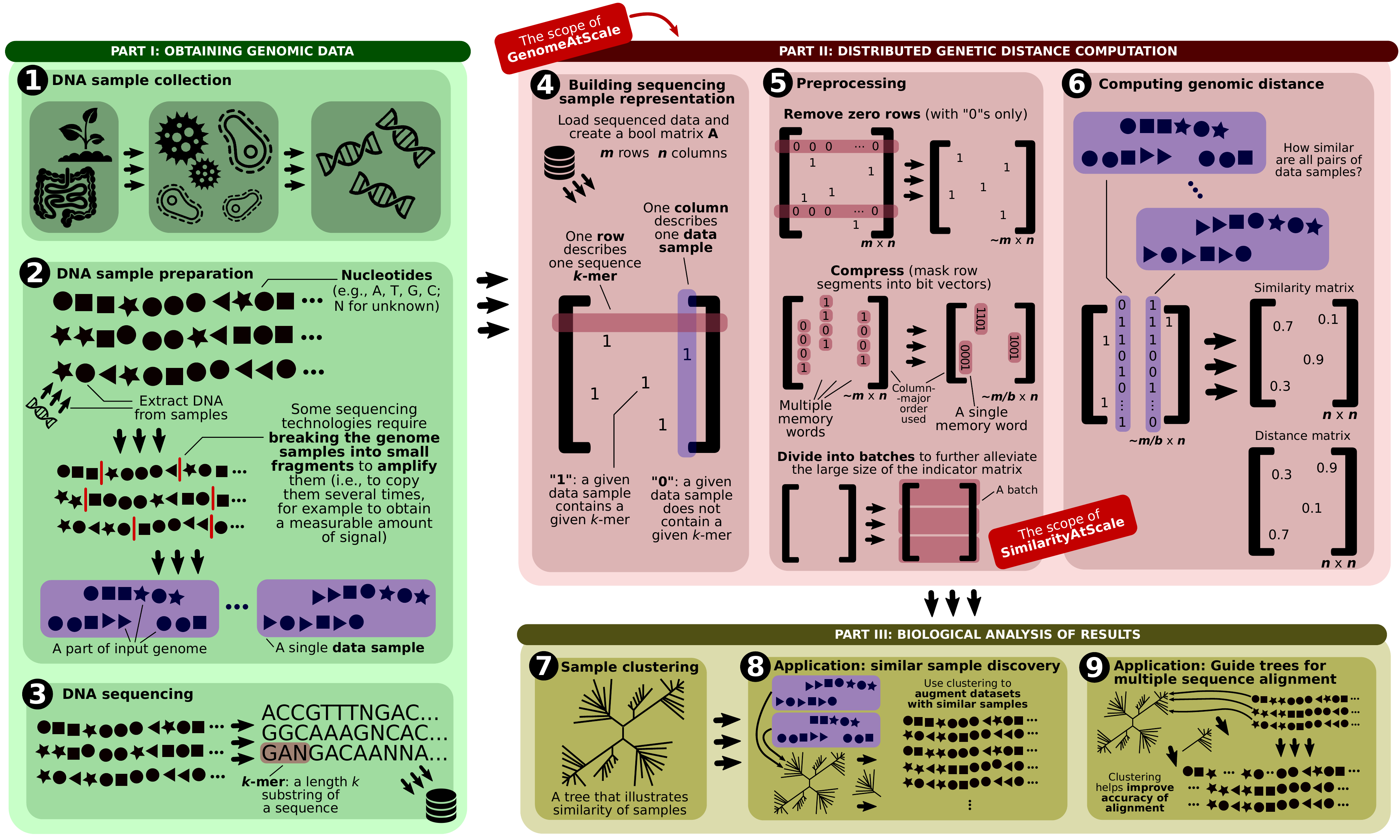}
\vspace{-1.5em}
\caption{\textbf{The scope of the SimilarityAtScale algorithm and GenomeAtScale tool within a metagenomics project.}
DNA is sequenced and preprocessed before being deposited into a database
%from different environment samples 
%or from a controlled setting
(\ding{182}--\ding{184}).
% The DNA is then fragmented into many small
% parts, amplified, and sequenced using a selected method such as shotgun
% sequencing~\cite{venter1998shotgun} (\ding{184}).
%The resulting reads are then deposited in a database as-is, or assembled into longer sequences beforehand.
Given sequence data from several samples, a binary matrix $\mat{A}$ is constructed indicating which $k$-mers are present in which samples (\ding{185}). The matrix is divided into batches and pair-wise Jaccard similarities are computed (\ding{186}--\ding{187}). The results may then be used for downstream genomics analysis (\ding{188}--\ding{190}).
%
%The results from the computed distance matrix can be used for various downstream genomics analysis tasks (\ding{188}--\ding{190}).
%This step is used to enable
%a fair treatment of underrepresented populations. 
%
%In the end, information in
%genomes is stored using a selected file format.
%
}
\vspace{-1.5em}
\label{fig:flow}
\end{figure*}

\section{Implementation}
\label{sec:genomeatscale}

We package the SimilarityAtScale algorithm as part of \textbf{GenomeAtScale}, a
tool for fast distributed genetic distance computation.
Figure~\ref{fig:flow} illustrates the integration of GenomeAtScale with general
genomics and metagenomics projects.  GenomeAtScale includes infrastructure to
produce files with a sorted numerical representation for each data sample.
Each processor is responsible for reading in a subset of these files, scanning
through one batch at a time.  Once the data is read-in, the SimilarityAtScale
implementation performs preprocessing and parallel sparse matrix
mutliplication.

To realize both preprocessing and sparse matrix multiplication, we use the
Cyclops library~\cite{solomonik2014massively}.  Cyclops is a distributed-memory
library that delivers routines for contraction and summation of sparse and
dense tensors.  The library also provides primitives for sparse input and
transformation of data.  Importantly, Cyclops enables the user to work with
distributed vectors/matrices/tensors with arbitrary fixed-size element
data-types, and to perform arbitrary elementwise operations on these
data-types.  This generality is supported via C++ templating, lambda functions,
and constructs for algebraic structures such as monoids and
semirings~\cite{Solomonik:2017:SBC:3126908.3126971,solomonik2015sparse}.
Cyclops relieves the user of having to manually determine the matrix data
distribution. Cyclops automatically distributes matrices over all processors
using a processor grid.  Each routine searches for an optimal processor grid
with respect to communication costs and any additional overheads, such as data
redistribution.

Listing~\ref{lst:alg} provides the pseudocode for our overall approach.  We
describe details of how preprocessing and similarity calculation are done with
Cyclops below.

\subsection{Implementation of Preprocessing}
\label{subsec:impl_prep}

After obtaining genome data (Part~I), we load the input sequence files (using
the established FASTA format~\cite{lipman1985rapid}) and construct a sparse
representation of the indicator matrix~$\mathbf{A}$.  The
\texttt{readFiles()} function then processes the input data, and writes into a
Cyclops sparse vector \textbf{f}, i.e., each $k$-mer is treated as an index to
update \textbf{f} with $1$.  To do this, we leverage the Cyclops
\texttt{write()} function, which collects arbitrary inputs from all processes,
combining them by accumulation, as specified by the algebraic structure
associated with the tensor.  We make use of the $(\max, \times)$ semiring for
\textbf{f} so that each vector entry is 1 if \emph{any} processor writes 1 to
it.

Our implementation then proceeds by collecting the sparse vector \textbf{f} on
all processors, and performing a local prefix sum to determine appropriate
nonzero rows for each data item.  This approach 
has been observed to be
most efficient for the
scale of problems that we consider in the experimental section.  The use of the
Cyclops \texttt{read()} function in place of replication would yield an
implementation that matches the algorithm description and communication cost.
The function \texttt{preprocessInput()} then proceeds to map the locally stored
entries to nonzero rows as prescribed by the prefix sum of the filter and to
apply the masking.  The distributed Cyclops matrix storing the batch of the
indicator matrix $\mat{\bar{A}}^{(l)}$ is then created by a call to
\texttt{write()} from each processor with the nonzero entries it generates.

\subsection{Implementation of Semiring Sparse Matrix Multiplication}
\label{subsec:impl_mult}

Given the generated sparse matrix $\mat{\bar{A}}^{(l)}$, which corresponds to
\textbf{A} in our pseudocode, the function \texttt{jaccardAccumulate()}
proceeds to compute the contribution to $\mat{B}$.  To do this with Cyclops, we
define a dense distributed matrix \textbf{B}, and use the Einstein summation
syntax provided by Cyclops to specify the matrix-multiplication.  The use of
the appropriate elementwise operation is specified via a Cyclops
\texttt{Kernel} construct, which accepts an elementwise function for
multiplication (for us \texttt{popcount}, which counts number of set bits via a
hardware-supported routine) and another function for addition, which in our
case is simply the addition of 64-bit integers.  The Einstein summation notation
with this kernel is used as follows
\begin{lstlisting}[numbers=none, basicstyle=\tt\scriptsize]
Jaccard_Kernel(|\textbf{A}|["ki"],|\textbf{A}|["kj"],|\textbf{B}|["ij"]);
\end{lstlisting}
Aside from this sparse matrix multiplication, it suffices to compute a column-wise
summation of the matrix \textbf{A}, which is done using similar Cyclops
constructs.

The resulting implementation of sparse matrix product is fully parallel,
and can leverage 3D sparse matrix multiplication algorithms. The
$\sqrt{p/c}\times\sqrt{p/c}\times c$ processor grid proposed in Section~\ref{sec:par_analysis}
can be delivered by Cyclops.  
%
\iffalse % REVISION
\sout{Consequently,
our use of Cyclops for parallel semiring sparse matrix multiplication should
achieve the sought-after communication cost and parallel scaling.}
%
\fi
%
\hl{Our results (almost ideal scaling) confirm that the desired
scaling is achieved for the considered parameters' spectrum.}
 
\ifrev
\marginpar{\vspace{-3em}\colorbox{orange}{\textbf{R1}}}
\fi

\begin{figure*}[t]
\vspace{-1em}
\centering
 \subfloat[\textbf{Kingsford dataset}, strong scaling.]{
  \includegraphics[width=0.32\textwidth]{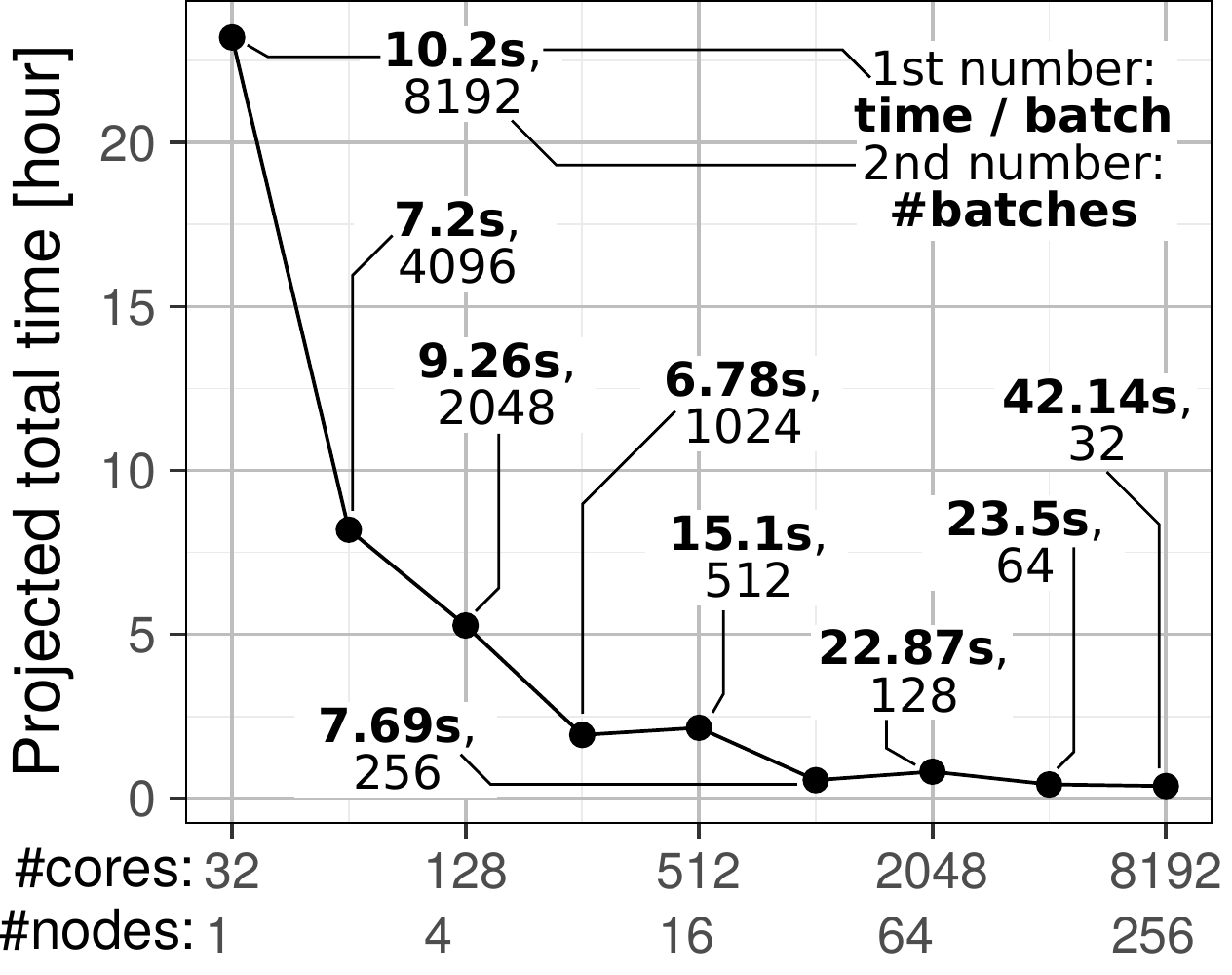}
  \label{fig:kingsford-strong}
 }%\hfill
 \subfloat[\textbf{BIGSI dataset}, strong scaling.]{
  \includegraphics[width=0.32\textwidth]{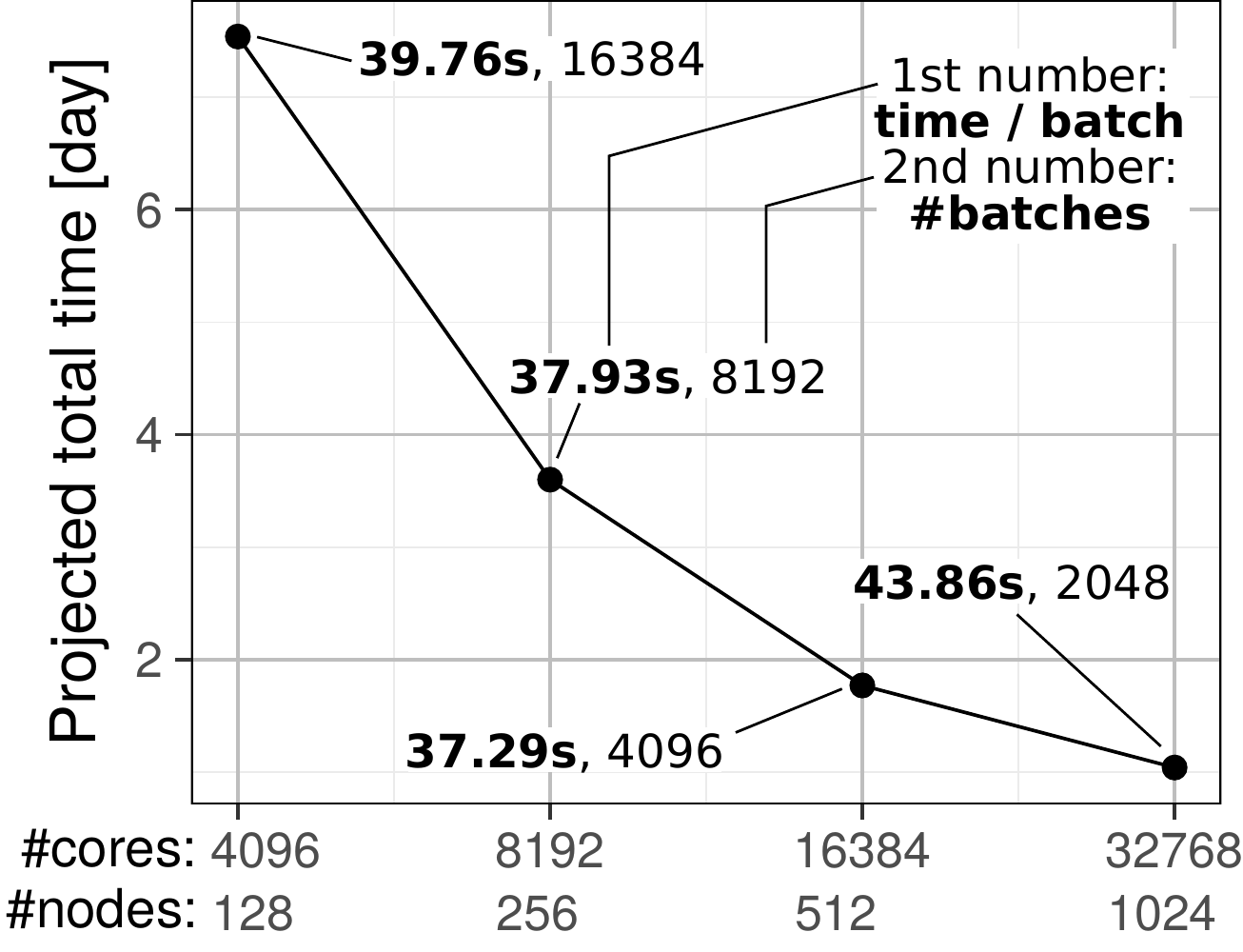}
  \label{fig:bigsi-strong}
 }%\hfill
 \subfloat[\textbf{Kingsford dataset}, sensitivity analysis.]{
  \includegraphics[width=0.32\textwidth]{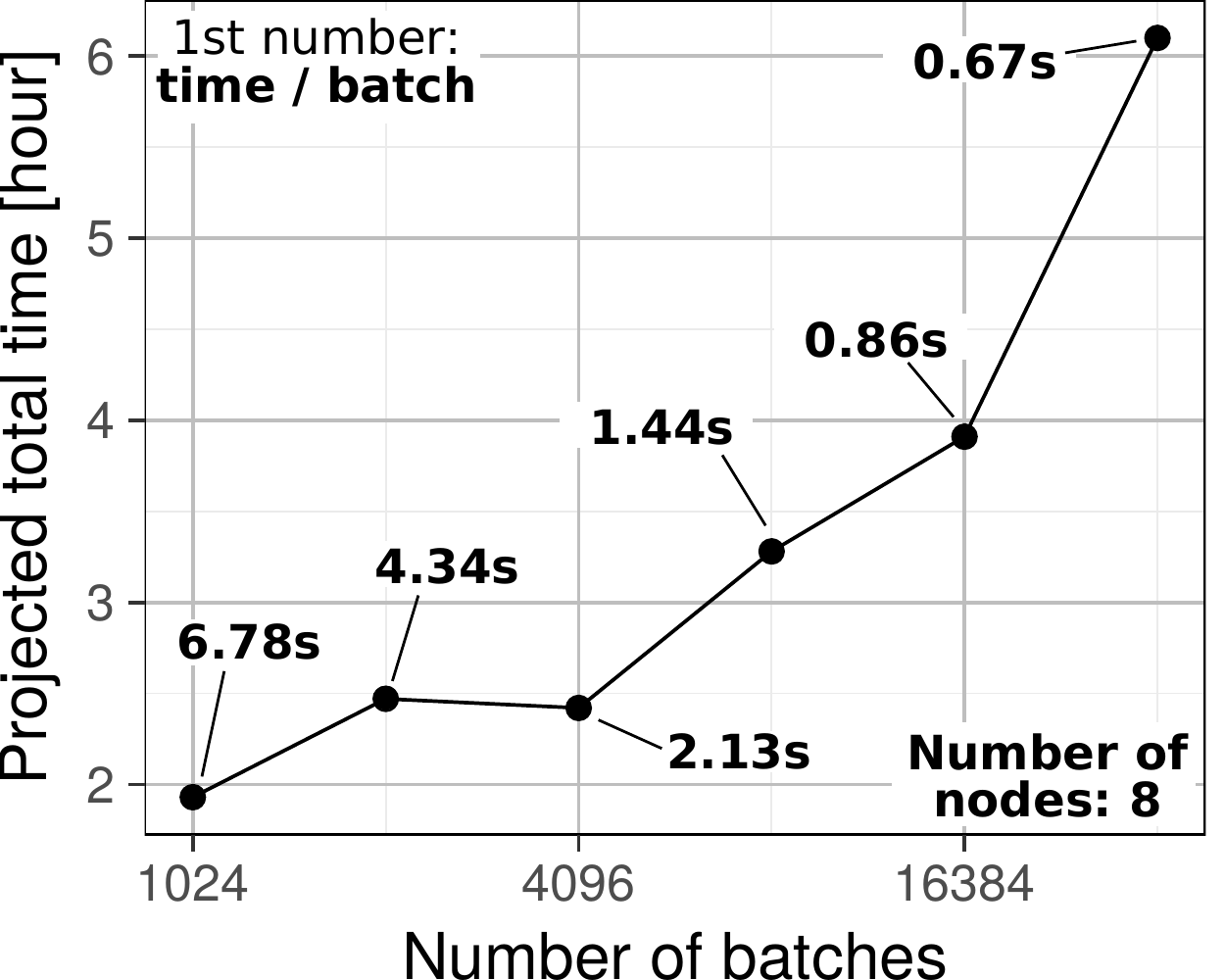}
  \label{fig:kingsford-batch}
 }\\
 \subfloat[\textbf{BIGSI dataset}, sensitivity analysis.]{
  \includegraphics[width=0.32\textwidth]{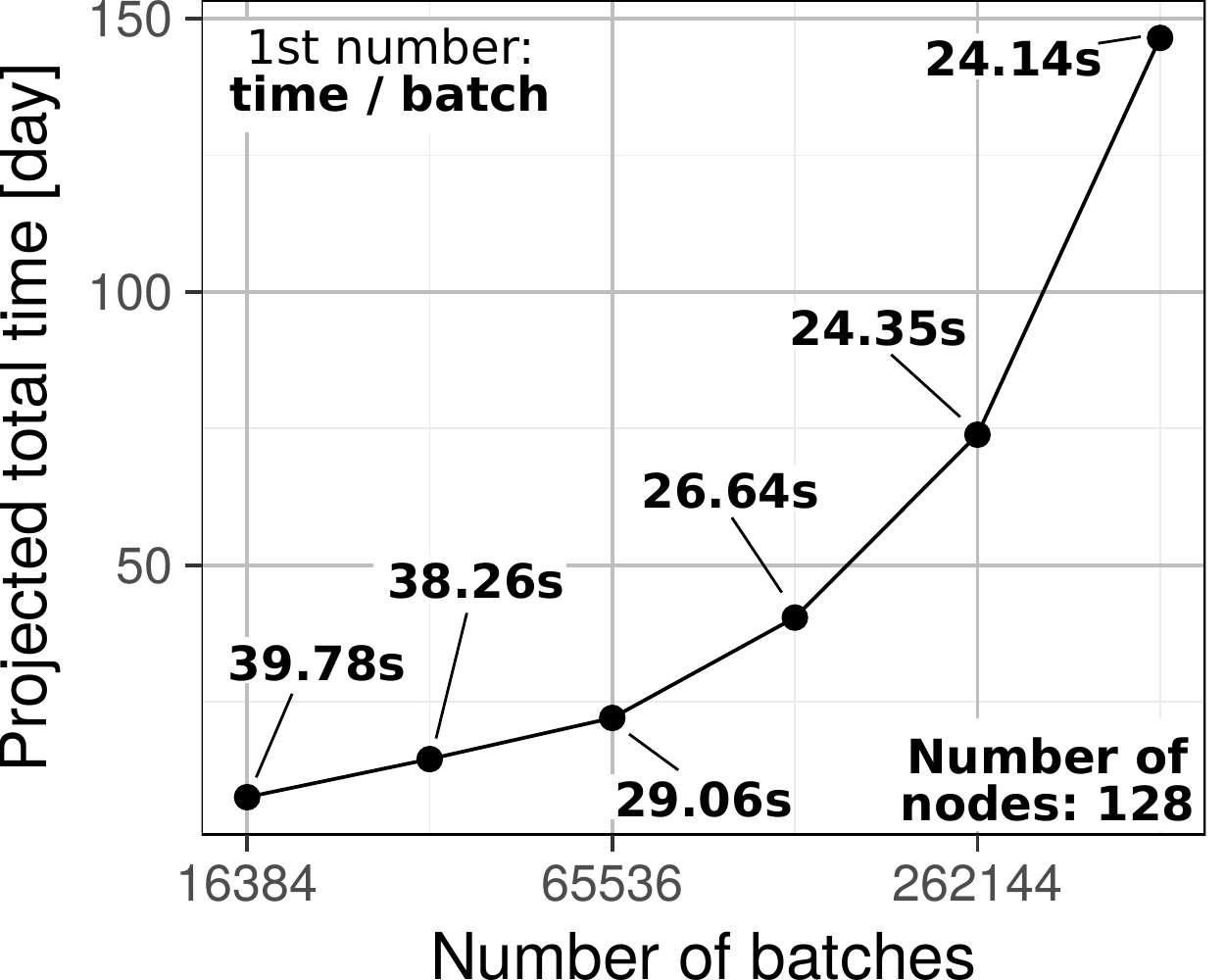}
  \label{fig:bigsi-batch}
 }%\hfill
 \subfloat[\textbf{Synthetic dataset}, strong scaling.]{
  \includegraphics[width=0.32\textwidth]{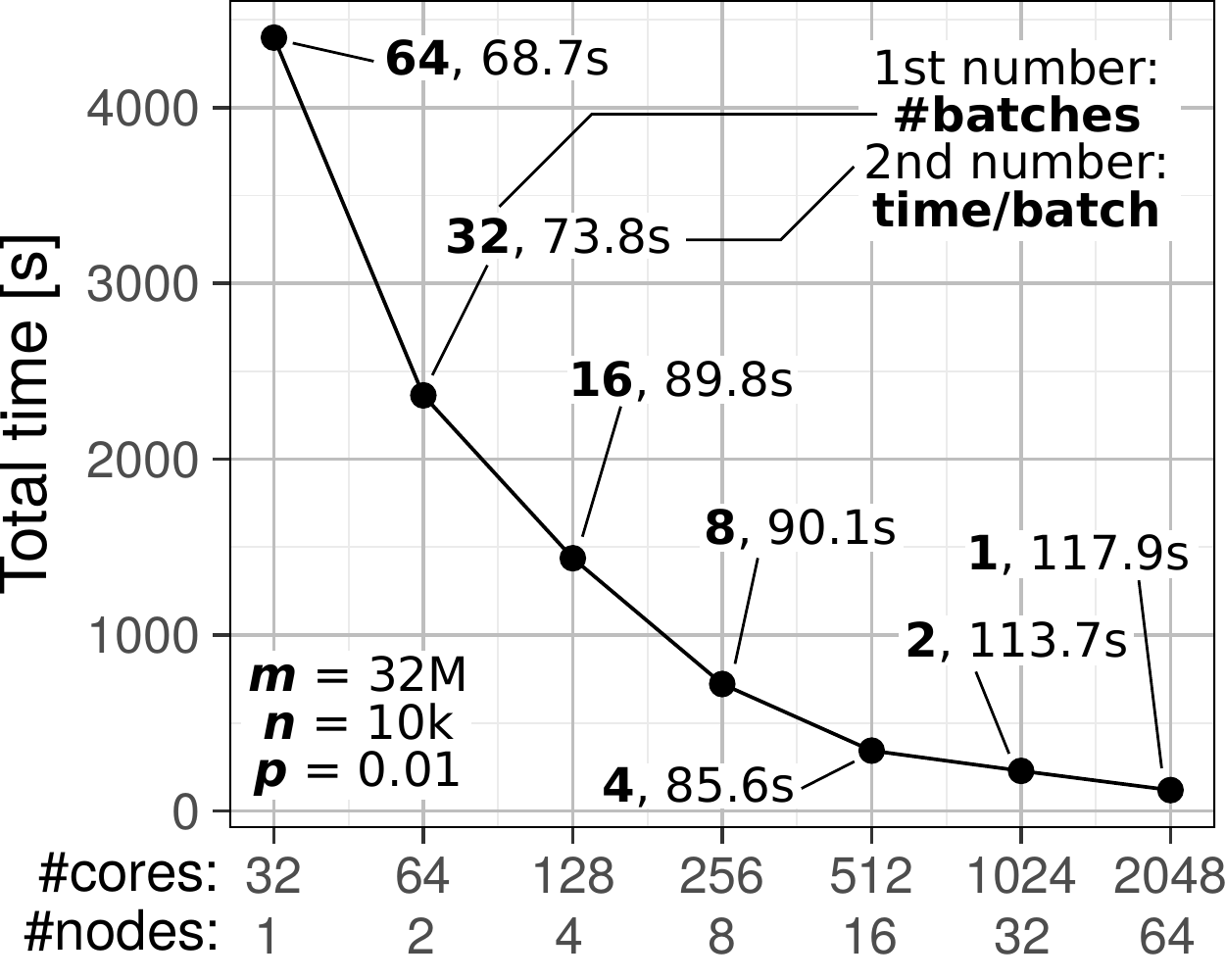}
  \label{fig:syn-strong}
 }%\hfill
 \subfloat[\textbf{Synthetic dataset}, weak scaling.]{
  \includegraphics[width=0.32\textwidth]{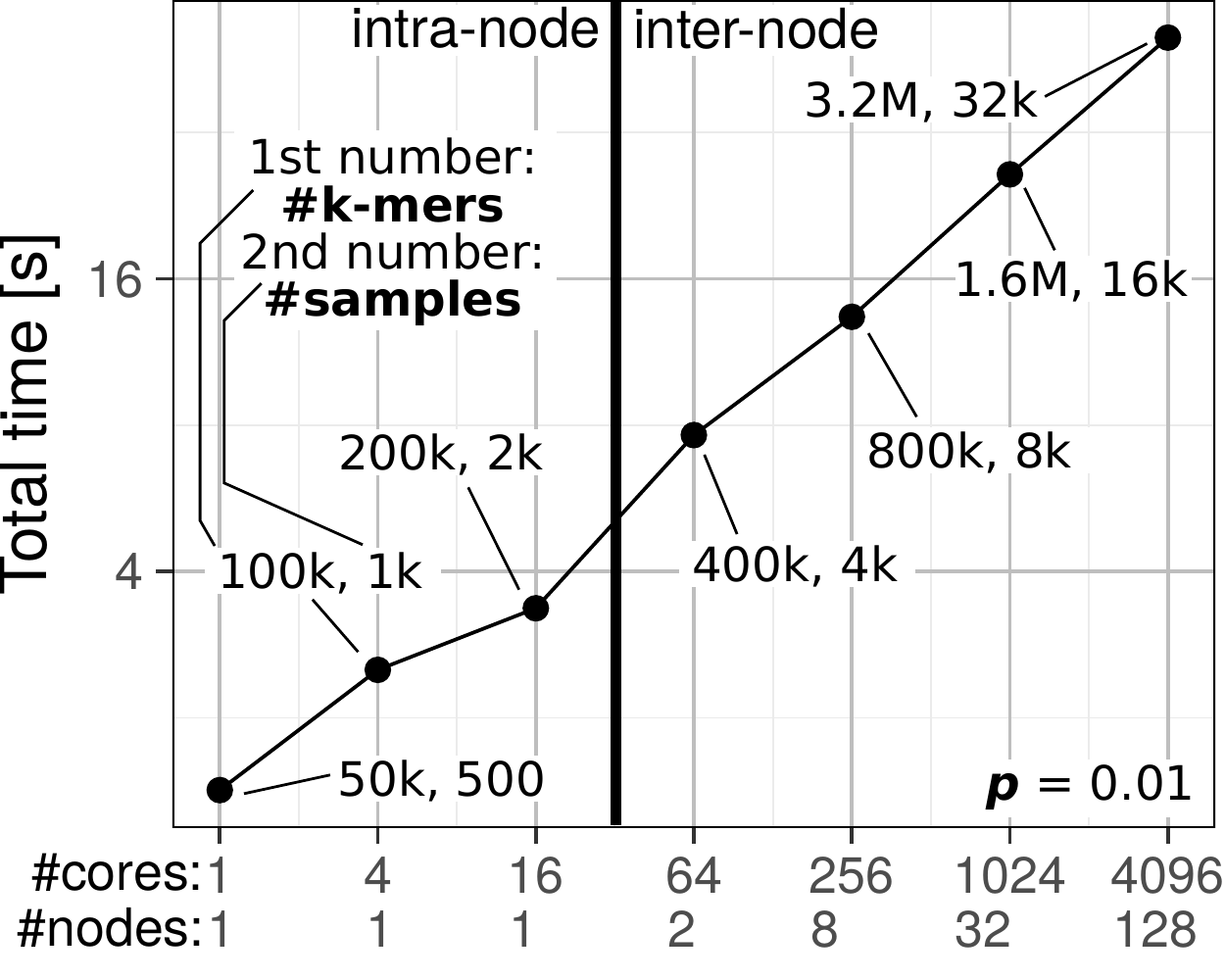}
  \label{fig:syn-weak}
 }
\vspace{-0.7em}
\caption{\textbf{Performance analysis of GenomeAtScale}. \hl{We calculate 95\% confidence intervals
for the reported mean values by assuming the batch times are normally distributed samples. \textbf{The derived confidence intervals
are very tight around the means, and we exclude them from the plot to ensure clarity.}
The confidence values for the BIGSI dataset are 0.12 (for 128 nodes), 0.16 (for 256 nodes), 0.38 (for 512 nodes),
and 0.40 (for 1024 nodes)}.}
\vspace{-1.5em}
\label{fig:eval-synthetic}
\end{figure*}

\section{Evaluation}
\label{sec:eval}

We now analyze the performance of our implementation of SimilarityAtScale for real and synthetic datasets.

\subsection{Methodology and Setup}

We first provide information that is required for interpretability and
reproducibility~\cite{hoefler2015scientific}.

\subsubsection{Experimental Setup}

We use the Stampede2 supercomputer. Each node has a Intel Xeon Phi 7250 CPU
(``Knights Landing'') with 68 cores, 96GB of DDR4 RAM, and 16GB of high-speed
on-chip MCDRAM memory (which operates as 16GB direct-mapped L3).
There is also 2KB of L1 data cache per core and 1MB of L2 per two-core tile.
There are 4,200 compute nodes in total. The network is a fat tree
with six core switches, with the 100 Gb/sec Intel Omni-Path architecture.

In our experiments, we consistently use 32 MPI processes per node. 
Using fewer processes per node enables larger batch sizes as our implementation
replicates the filter vector on each processor.
We also find that this configuration outperforms those with 
64 processes per node for representative experiments, as the 
on-node computational kernels are generally memory-bandwidth bound.
%ensures performance benefits (using even fewer processes could
%negatively impact performance due to less parallelism).  We verify this by
%running representative experiments for 64 processes.
\hl{To maximize fair evaluation, we include the I/O time (loading data from disk)
in the reported runtimes (the I/O time is $\approx$1\% of the total runtime).
}

\ifrev
\marginpar{\vspace{-2em}\colorbox{orange}{\textbf{R3}}}
\fi

\iffalse
%
We use Blue Waters, a Cray XE6 supercomputer. Each Blue Waters XE node has two
16-core AMD 6276 Interlagos sockets; there are 22,500 nodes in total. The
network is a Cray Gemini torus. 
%
\fi

\subsubsection{Considered Real Datasets}\label{ref:datasets}
We evaluate our design on datasets of differing sequence variability to
demonstrate its scalability in different settings. As a \textbf{low-variability
set}, we use the public BBB/Kingsford dataset consisting of 2,580 RNASeq
experiments sequenced from human blood, brain, and breast
samples~\cite{solomon2016fast} with sequencing reads of length at least 20.
We consider
all such experiments that were publicly available at the time of study.  The raw
sequences were preprocessed to remove rare (considered noise) $k$-mers. Minimum
$k$-mer count thresholds were set based on the total sizes of the raw
sequencing read sets~\cite{solomon2016fast}. We used the $k$-mer size of $19$ (unlike
the value of $20$ used in \cite{solomon2016fast}) to avoid the possibility of $k$-mers
being equal to their reverse complements.
As a \textbf{high-variability set}, we use all bacterial and viral
whole-genome sequencing data used in the BIGSI
database~\cite{bradley2019ultrafast}, representing almost every such experiment
available as of its release, totaling 446,506 samples (composed overwhelmingly of Illumina short-read sequencing experiments). In the
same fashion as the BIGSI, these data were preprocessed by
considering longer contiguous stretches of $k$-mers to determine $k$-mer count
thresholds~\cite{bradley2019ultrafast}. We used the same $k$-mer size as the BIGSI paper ($k = 31$).

The considered real datasets enable analyzing the performance of our schemes
for different data sparsities. 
  %
%Specifically, after filtering, 
%\new{We considered versions of the two datasets that were preprocessed outside of GenomeAtScale, as well as raw unfiltered versions of the datasets.
%Prior to filtering, the Kingsford dataset yields an indicator matrix $\mat{A}$ with a density of roughly $1.5 \cdot 10^{-4}$ and the BIGSI dataset has a density of roughly $4 \cdot 10^{-12}$.
%The filtered versions of the datsets, which omit noise and remove many of the zero rows, yield a density of nonzeros in the indicator matrix $\mat{A}$ of roughly $10^{-1}$ for the Kingsford dataset and $10^{-5}$ for the BIGSI dataset.}
%
The indicator matrix~$\mat{A}$ in the Kingsford dataset has a density of $\approx$$1.5 \cdot 10^{-4}$, and in the BIGSI dataset its density is $\approx$$4 \cdot 10^{-12}$.
All input data is provided in the FASTA format~\cite{lipman1985rapid}.

\subsubsection{Considered Synthetic Datasets}

We also use synthetic datasets where each element of the indicator
matrix~$\mat{A}$ is present with a specified probability~$p$ (which corresponds to density),
independently for all elements. This enables a systematic analysis of the
impact of data sparsity on the performance of our schemes.

\subsubsection{Considered Scaling Scenarios}

To illustrate the versatility of our design, we consider (1) strong
scaling for a small dataset (fixed indicator matrix ($\mat{A}$) size, increasing batch size
and core count), (2) strong scaling for a large dataset (same as above), 
%
%(3) strong scaling for a single batch, 
%
(3) weak scaling (increasing the $\mat{A}$ size with core count, increasing batch size with core count), 
%
%\raghu{matrix size here is the output matrix, and batch size the input matrix. do we need to explain it a bit?} 
%
(4) batch
size sensitivity (fixed node count, increasing batch size).
%
%, (6) sparsity sensitivity (fixed indicator matrix size, varying sparsity).

\subsection{Performance Analysis for Real Data}

The results for the Kingsford and the BIGSI datasets are presented in
Figure~\ref{fig:kingsford-strong} and~\ref{fig:bigsi-strong}, respectively.
The BIGSI dataset as noted has $n=446,506$ columns. This requires us to
distribute three matrices $\mat{A}$, $\mat{B}$, and $\mat{C}$ among the processes
for the similarity calculation. 
%In 64 nodes, we run out of memory for the same.
We find it necessary to use 64 nodes to have enough memory to store these matrices.
  Hence, we report performance numbers for 128, 256, 512, and 1024 nodes. 
  As we double the number of nodes, we also double the batch size,
  utilizing all available memory. 
  We also use not more than 256 nodes for preprocessing, i.e., the sparse vector is constructed using not more than 256 nodes, but is used by all the participating nodes in the later stages of the pipeline. We find less variability in performance with this variant.
%Our timing are based on the average execution time out of eleven batches.

In Figure~\ref{fig:bigsi-strong}, we show the average batch time (averaged
across eight batches, not considering the first three batches to account for
startup cost). Per batch time across nodes, remains the same (the batch size
though is doubled according to the node size). The y-axis shows the predicted
completion time for running the entire dataset to calculate the similarity
matrix. We note that we are able to calculate the similarity matrix for BIGSI
benchmark in a day (24.95 hours) using 1024 nodes.
We note that despite high-variability of density across different columns in the BIGSI dataset, SimilarityAtScale achieves a good parallel efficiency even when using 32K cores.

Similarly, in Figure~\ref{fig:kingsford-strong}, we show the results for the
Kingsford dataset which is denser than BIGSI. The performance behavior observed for this smaller dataset is a bit less consistent.
We observe both superscalar speed-ups and some slow-downs when increasing node count.
On 32 nodes, we note a sweet-spot, achieving a 42.2$\times$ speed-up relative to the single node performance.
Thus, we are able to construct the similarity matrix in less than an hour using 32 nodes. 
For larger node counts, the number of MPI processes (2048, 4096, 8192) starts to exceed the number of columns in the matrix (2,580), leading to load imbalance and deteriorating performance.
%for completion does not follow a smooth curve. We note that the batch size used
%  in each of the runs is different i.e., batch size is doubled with the
%  doubling in the node count. And any imbalance in the distribution of 1's
%  ($k$-mers) might affect the performance for that batch.  \raghu{i.e., a 512 node processing a
%  bigger batch when compared to 256 is not just handling double the data size,
%  but might be handling more than double the number of 1's in the input matrix.
%  Will refine this statement.} 
%However,

\ifrev
\marginpar{\vspace{3em}\colorbox{orange}{\textbf{R1}}}
\fi

\hl{
To verify the projected execution times, we 
fully process Kingsford for 128 nodes and 64 batches (we cannot
derive data for \emph{all} parameter values due to budget constraints).
The total runtime takes 0.38h, the corresponding projection is 0.42h.
}

\ifrev
\marginpar{\vspace{-22em}\colorbox{orange}{\textbf{R1}}}
\fi

In Figures~\ref{fig:kingsford-batch} and \ref{fig:bigsi-batch}, we show the sensitivity of the datasets for the size of the batches. 
For both datasets we observe a general trend that the execution time does not scale with batch size, despite the work scaling linearly with batch size.
This behavior is expected, as a larger batch size has a lesser overhead in synchronization/latency and bandwidth costs, enabling a higher rate of performance.
%In Kingsford, where the dataset has a density of roughly $1.5 \cdot 10^{-4}$, time per batch reduces with the decrease in batch size. Though the timer per batch reduces with decrease in batch size even in BIGSI, there is no linear correlation between the time reduction, and the decrease in batch size. The time per batch roughly remains the same for both the resultant batch sizes when run with 16384 and 32768 batches, respectively. BIGSI dataset is much more sparser with its density being roughly $4 \cdot 10^{-12}$.
%\raghu{should we say CTF handles sparsity in a way that it scales?}
Thus, in both the datasets the overall projected time for the similarity matrix calculation reduces with the increase in batch size.

\subsection{Performance Analysis for Synthetic Data}
In Figure~\ref{fig:syn-strong}, we present the strong scaling results (with the increasing batch size) for synthetic data. 
The total time decreases in proportion to the node count, although the time per batch slightly increases, yielding good overall parallel efficiency, as predicted in our theoretical analysis. 
In Figure~\ref{fig:syn-weak}, we show weak scaling by increasing the $\mat{A}$ matrix size \emph{and} the batch size with the core count.
In this weak scaling regime, the amount of work per processor is increasing with the node count.
From 1 core to 4096 cores, the amount of work per processor increases by 64$\times$, while the execution time increases by 35.3$\times$, corresponding to a 1.81$\times$ efficiency improvement.

%The results for the synthetic datasets are presented in
%Figure~\ref{fig:syn-strong} and~\ref{fig:syn-weak}, respectively.

%
%\maciej{TODO: discussion}

\begin{figure}[h]
\vspace{-1em}
\centering
  \includegraphics[width=0.32\textwidth]{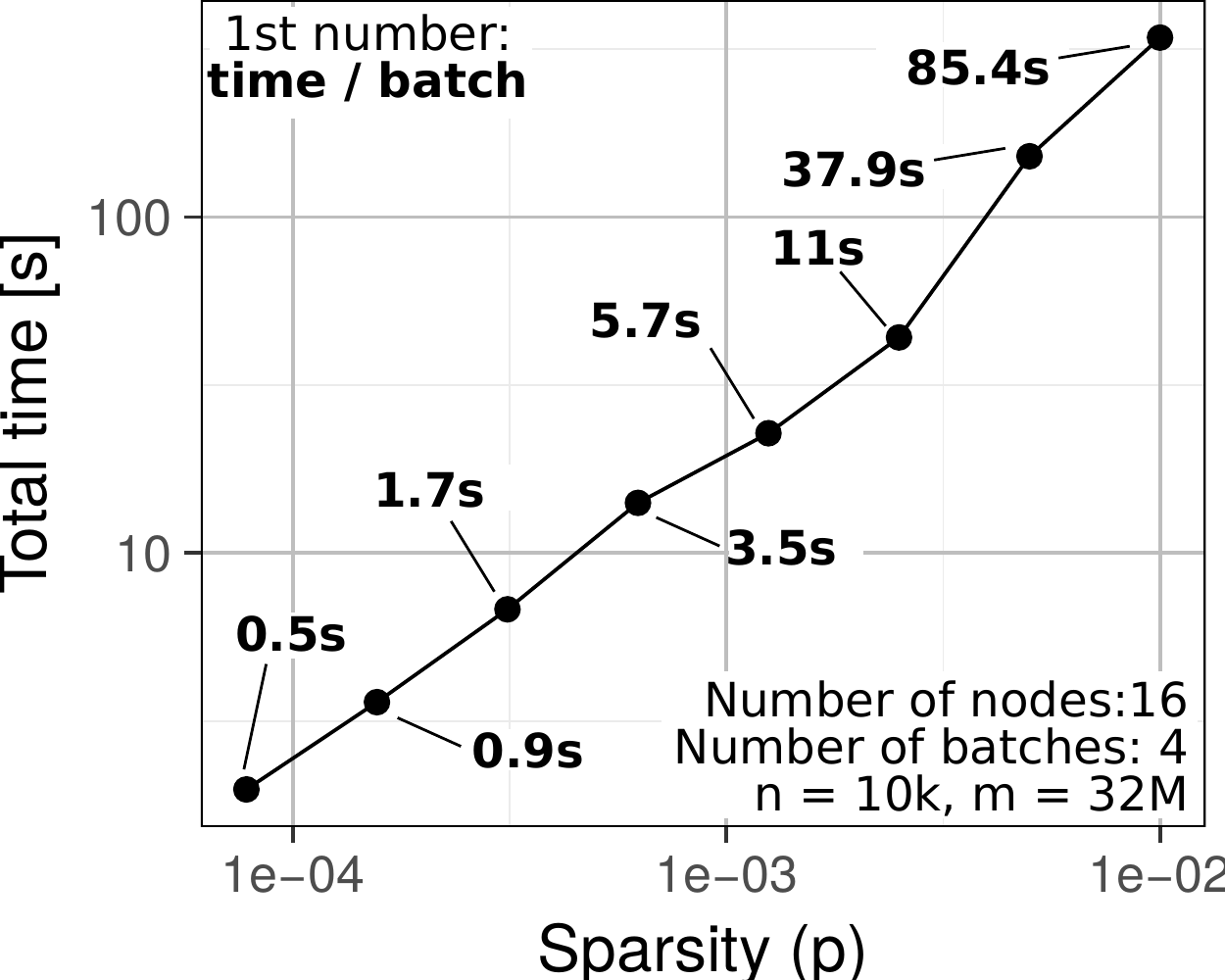}
%\vspace{-0.8em}
%
\caption{\hl{The impact of data sparsity on the performance of GenomeAtScale.}}
\vspace{-0.75em}
\label{fig:eval-synthetic-p}
\end{figure}

\hl{
We also show how the performance for synthetic datasets changes with data sparsity
expressed with the probability~$p$ of the occurrence of a particular $k$-mer.
The results are in Figure~}\colorbox{white}{\ref{fig:eval-synthetic-p}}\hl{.
We enable nearly ideal scaling of the \emph{total runtime}
with the decreasing data sparsity (i.e., with more data to process).
}

\ifrev
\marginpar{\vspace{-3em}\colorbox{orange}{\textbf{R1}}}

\marginpar{\vspace{4em}\colorbox{orange}{\textbf{R1}}}
\fi

\iffalse
%
\begin{figure}[h]
\centering
%\vspace{-1em}
 \includegraphics[width=0.35\textwidth]{kingsford_strong___e.pdf}
%
\caption{Performance of GenomeAtScale for the \textbf{Kingsford dataset} (strong scaling).}
%
%\vspace{-0.5em}
\label{fig:kingsford-strong}
\end{figure}

\begin{figure}[h]
\centering
%\vspace{-1em}
 \includegraphics[width=0.35\textwidth]{bigsi_strong___e.pdf}
%
\caption{Performance of GenomeAtScale for the \textbf{BIGSI dataset} (strong scaling).}
%
%\vspace{-0.5em}
\label{fig:bigsi-strong}
\end{figure}

\begin{figure}[h]
\centering
%\vspace{-1em}
 \includegraphics[width=0.35\textwidth]{kingsford_batch___e.pdf}
%
\caption{Sensitivity analysis of GenomeAtScale for the \textbf{Kingsford dataset}.}
%
%\vspace{-0.5em}
\label{fig:kingsford-batch}
\end{figure}

\begin{figure}[h]
\centering
%\vspace{-1em}
 \includegraphics[width=0.35\textwidth]{bigsi_batch___e.pdf}
%
\caption{Sensitivity analysis of GenomeAtScale for the \textbf{BIGSI dataset}.}
%
%\vspace{-0.5em}
\label{fig:bigsi-batch}
\end{figure}

\begin{figure}[h]
\centering
%\vspace{-1em}
 \includegraphics[width=0.35\textwidth]{synthetic_strong___e.pdf}
%
\caption{Performance analysis of GenomeAtScale for \textbf{synthetic datasets} (strong scaling, $m=32$M, $n=10$k,
$p=0.01$).}
%
%\vspace{-0.5em}
\label{fig:syn-strong}
\end{figure}

\begin{figure}[h]
\centering
%\vspace{-1em}
 \includegraphics[width=0.35\textwidth]{synthetic_weak___e.pdf}
%
\caption{Performance of GenomeAtScale for \textbf{synthetic datasets} (weak scaling, 
$p=0.01$).}
%
%\vspace{-0.5em}
\label{fig:syn-weak}
\end{figure}
%
\fi

\subsection{\hl{Impact from Fast Cache}}

\ifrev
\marginpar{\vspace{3em}\colorbox{orange}{\textbf{R3}}}
\fi

\hl{
We also test our design \emph{without using MCDRAM as L3 cache},
but instead as an additional memory storage.
The resulting performance patterns are negligibly worse than the ones in which
MCDRAM serves as L3. For example, time per batch with MCDRAM as
L3 for Kingsford dataset on 4 nodes and 32 nodes is 9.26s and 7.69s,
respectively. Then, without MCDRAM L3 cache, it is 9.33s and 8.01s, respectively.
}

\vspace{-0.75em}
\section{Discussion and Related Work}
\label{sec:discussion}
\vspace{-0.3em}

% We now discuss different related aspects of our designs.

% \subsection{SimilarityAtScale with MapReduce}

The algebraic formulation of our schemes is generic and can be implemented with
other frameworks such as CombBLAS~\cite{bulucc2011combinatorial,
kepner2016mathematical}. We selected Cyclops which is -- to the best of our
knowledge -- the only library with high-performance routines where
\emph{the input matrices are sparse but the outcome of the matrix-matrix
multiplication is dense}.
CombBLAS targets primarily graph processing and, to the best of our knowledge,
does not provide a fast implementation of matrix-matrix product with a dense
output.  Thus, CombBLAS would result in suboptimal performance when combined
with SimilarityAtScale. 
One could also use MapReduce~\cite{dean2008mapreduce} and engines such as
Spark~\cite{zaharia2010spark} or Hadoop~\cite{white2012hadoop}.
Yet, they are communication-intensive and their limited expressiveness often
necessitates multiple communication rounds, \emph{resulting in inherent
overheads when compared to communication-avoiding and expressive algebraic
routines provided by Cyclops}.

\iffalse

A recent closely related work, BELLA~\cite{bella2019} leverages multiplication
of sparse matrices for the computation of pairwise overlaps between long reads
as a part of genome assembly via Overlap Layout Consensus
(OLC)~\cite{chu2016innovations}, and has been extended to support parallel
execution~\cite{ellis2019dibella}. Our work, on the other hand, targets whole
genome comparison, computing distances between entire read sets. Consequently,
our parallelization strategy is very different. More details comparing these
two approaches are in Section~\ref{sec:discussion}.

\fi

A recent closely related work, BELLA~\cite{bella2019, ellis2019dibella}, is the
first to express similarity calculation for genetic sequences as
sparse matrix-matrix multiplication. BELLA leverages parallel multiplication of
sparse matrices for the computation of pairwise overlaps between long reads as
a part of genome assembly~\cite{georganas2014parallel, chapman2015whole,
georganas2015hipmer, georganas201718, georganas2017merbench,
ellis2017performance, georganas2018extreme} via Overlap Layout Consensus
(OLC)~\cite{chu2016innovations}.
BELLA constructs an overlap graph of reads in a read set while
we target whole genome comparison, computing distances between entire
read sets.

BELLA performs overlap detection using a sparse matrix multiplication $\mat
A\mat{A}^T$, where $k$-mers correspond to columns of $\mat{A}$, and each row
corresponds to a read.  The data samples in BELLA are individual reads from the
same read set (e.g., organism), while we consider all $k$-mers in a read set
(e.g., corresponding to a whole genome) to constitute a single sample.  In our
context, $\mat{A}$ has many more nonzeros and more potential variability in
density among columns.  Further, the output Jaccard matrix is generally dense,
while for BELLA the overlap is frequently zero.  This fact has also motivated a
specialized parallelization of the overlap and alignment calculations for
BELLA~\cite{ellis2019dibella}, which are not based on parallel sparse matrix
multiplication. Additionally, SimilarityAtScale employs parallel methods for
compressing the input data, and the output matrix is computed in batches with
the input matrix also constructed in batches, which are not necessary in the
context of long reads in BELLA.

\section{Conclusion}

We introduce \textbf{SimilarityAtScale}, the first high-performance distributed
algorithm for computing the Jaccard 
similarity, which is widely used in data analytics.
SimilarityAtScale is based on an
algebraic formulation that uses (1) provably
communication-efficient linear algebra routines, (2) compression based on bitmasking, (3) batched
computation to alleviate large input sizes, and (4) theoretical analysis that
illustrates scalability in communication cost and parallel efficiency.
The result is a generic high-performance algorithm that can be applied to any
problem in data analytics that relies on Jaccard measures. To facilitate the
utilization of SimilarityAtScale in different domains, we provide a
comprehensive overview of problems that could be accelerated and scaled with
our design.

We then use SimilarityAtScale as a backend to develop \textbf{GenomeAtScale},
the first tool for accurate large-scale calculations of distances between
high-throughput whole-genome sequencing samples on distributed-memory systems. To foster DNA research, we
use real established datasets in our evaluation, showing that -- for
example -- GenomeAtScale enables analyzing all the bacterial and viral
whole-genome sequencing data used in the BIGSI database in less than a day. We
maintain compatibility with standard bioinformatics data formats, enabling
seamless integration of GenomeAtScale with existing biological analysis
pipelines. We conduct largest-scale exact computations of
Jaccard genetic distances so far. Our publicly available design and implementation can
be used to foster further research into DNA and general data analysis.

\section*{Acknowledgment}

This work used the Extreme Science and Engineering Discovery Environment (XSEDE), which is supported by National Science Foundation grant number ACI-1548562.
We used XSEDE to employ Stampede2 at the Texas Advanced Computing Center (TACC) through allocation TG-CCR180006.
We thank Andre Kahles and Grzegorz Kwasniewski for the discussions in the initial phase of the work.

{
  %\footnotesize
  %\bibliographystyle{ACM-Reference-Format}
  \bibliographystyle{abbrv}
  %\bibliography{sample-bibliography} 
  \bibliography{references}
}

%{
%  \printbibliography
%}

\end{document}